\RequirePackage{ifpdf}
\documentclass[12pt,letterpaper,bibliography=totoc,
    listof=totoc,
    final]{JHEP3}

\usepackage{amscd,amsmath,amssymb,amsfonts,xspace,mathrsfs,caption}
\usepackage[utf8]{inputenc}
\usepackage{graphicx}

 \usepackage{enumerate}
\newcommand{\be}{\begin{equation}}
\newcommand{\ee}{\end{equation}}
\newcommand{\bes}{\begin{equation*}}
\newcommand{\ees}{\end{equation*}}
\newcommand{\CA}{\mathcal{A}}
\newcommand{\CB}{\mathcal{B}}

\newcommand{\cD}{\mathcal{D}}

\newcommand{\CF}{\mathcal{F}}

\newcommand{\CH}{\mathcal{H}}
\newcommand{\CI}{\mathcal{I}}

\newcommand{\CK}{\mathcal{K}}
\newcommand{\CL}{\mathcal{L}}

\newcommand{\CN}{\mathcal{N}}
\newcommand{\CO}{\mathcal{O}}

\newcommand{\CQ}{\mathcal{Q}}
\newcommand{\CR}{\mathcal{R}}
\newcommand{\CS}{\mathcal{S}}
\newcommand{\CT}{\mathcal{T}}

\newcommand{\BR}{\mathbb{R}}

\newcommand{\BC}{\mathbb{C}}
\newcommand{\BH}{\mathbb{H}}

\newcommand{\BZ}{\mathbb{Z}}
\newcommand{\BQ}{\mathbb{Q}}

\newcommand{\bra}{\langle}
\newcommand{\ket}{\rangle}

\newcommand{\bfmu}{{\boldsymbol \mu}}

\newcommand{\bfx}{{\boldsymbol x}}
\newcommand{\bfy}{{\boldsymbol y}}

\newcommand{\bfk}{{\boldsymbol k}}

\newcommand{\bfz}{{\boldsymbol z}}

\newcommand{\TB}{\bar{\tau}}

\title{Renormalization and BRST symmetry in Donaldson-Witten theory}

\author{Georgios Korpas$^{1,2}$, Jan Manschot$^{1,2}$, Gregory W. Moore$^3$ and Iurii Nidaiev$^3$\\
{\it $^1$ School of Mathematics, Trinity College, Dublin 2, Ireland}\\
{\it $^2$ Hamilton Mathematical Institute, Trinity College, Dublin 2, Ireland}\\
{\it $^3$ NHETC and Department of Physics and Astronomy, Rutgers University, 126 Frelinghuysen Rd., Piscataway NJ 08855, USA}
\vspace*{2mm}\\ {\tt e-mail:
\email{george.korpas@maths.tcd.ie}, 
\email{manschot@maths.tcd.ie}, \email{gwmoore@physics.rutgers.edu}, \email{iurii.nidaiev@gmail.com}}
}

\abstract{The presence of a BRST symmetry in topologically twisted
  gauge theories makes a precise analysis of these
  theories feasible. While the global BRST symmetry suggests that
  correlation functions of BRST exact observables vanish, this
  decoupling might be obstructed due to a contribution from the boundary
  of field space. Motivated by divergent BRST exact observables on the Coulomb branch of
  Donaldson-Witten theory, we put forward a new prescription for the
  renormalization of correlation functions on the Coulomb branch. This
 renormalization is based on the  relation between Coulomb branch integrals and
  integrals over a modular fundamental domain, and establishes that BRST exact
  observables indeed decouple in Donaldson-Witten theory.\\ }

\preprint{TCD-19-1}

\begin{document}

%%%%%%%%%%%%%%%%%%%%%%%%%%%%%%%%%%%%%%%%%%%%%%
%%%%%%%%%%%%%%%%%%%%%%%%%%%%%%%%%%%%%%%%%%%%%%
%%%%%%%%%%%%%%%%%%%%%%%%%%%%%%%%%%%%%%%%%%%%%%
%%%%%%%%%%%%%%%%%%%%%%%%%%%%%%%%%%%%%%%%%%%%%%
%%%%%%%%%%%%%%%%%%%%%%%%%%%%%%%%%%%%%%%%%%%%%%
\section{Introduction}
Topological twists of supersymmetric quantum field theories have been of immense importance
in the last thirty years in both physics and mathematics. Such
theories are sometimes referred to as cohomological field theories
(CohFTs), and have provided the foundation for the physical
formulation of mathematically defined invariants, such as Donaldson invariants \cite{Witten:1988ze}
and Gromov-Witten invariants \cite{witten1988}. More recently,
they have played a prominent role in the study of the geometric Langlands
program \cite{Kapustin:2006pk}, and the evaluation of central charges
in superconformal theories \cite{Shapere:2008zf}.

Let us briefly recall the main principles of topological gauge
theories on a Riemannian four-manifold $M$ with metric $g$
\cite{Witten:1988ze, Witten:1990bs, Birmingham:1991ty, Cordes:1994fc, Moore:1997pc, LoNeSha}. One of their key properties is that they contain
a scalar fermionic BRST operator\footnote{It is possible that
  topological field theories contain more than one such operator as in ``balanced topological field
  theories" \cite{Dijkgraaf:1996tz, Blau:1991bn} of which   Vafa-Witten theory \cite{Vafa:1994tf} is an example.}, $\mathcal{Q}$, such that
$$\mathcal{Q}^2=0\,\, \rm{modulo\,\,gauge\,\,transformations}.$$
This operator divides observables
of the theory in three sets: i) those $\CO$ for
which\footnote{Throughout the text we use the brackets $\{,\}$ for both commutators and
  anti-commutators depending on the parity of $\CO$.} $\{\CQ,\CO\}\neq
0$, ii)
the $\CQ$-commutators (or $\CQ$-exact) ones, which can be expressed as $\CO=\{\CQ,\CO'\}$ for some
$\CO'$, and iii) the $\CQ$-closed ones, which satisfy $\{\CQ,\CO\}=0$ without
being $\CQ$-exact. An important example of a $\CQ$-exact operator in
such theories, is the variation of the action $\CS$ with respect to
the metric $g$. For a suitable $W_{\mu\nu}$, we can express it as
\be \label{deltagL} \delta_g \mathcal{S}(X)=\tfrac{1}{2} \int_M \sqrt{g}\, g^{\mu\nu} \{ \CQ,W_{\mu\nu} \},\ee
where $X$ is a short hand for the collection of fields of the
theory.

The path integral measure $\cD X$ and the action $\CS(X)$ are
both invariant under the global symmetry $\CQ$. The Ward-Takahashi identity
for this symmetry then suggests that the vev of a gauge
invariant $\CQ$-exact operator vanishes,
\be
\label{QV0}
\left< \{{\CQ},\CO\}\, \right>=\int [\mathcal{D}X]\,\{{\CQ},\CO\}\,e^{-\mathcal{S}(X)}=0,
\ee
or equivalently,
\be
\label{QV1}
\sum_{i} \langle \{{\CQ},\CO_i\}\, \prod_{j\neq i} \CO_j \rangle=0.
\ee

Moreover, $\CQ$-exact observables decouple from $\CQ$-closed
observables, since
\be
\label{CQdecoupling}
 \langle \{\CQ,\CO'\} \prod_{j}\CO_j\rangle=0,\qquad \mathrm{if}\,\,\{\CQ,\CO_j\}=0\,\,\mathrm{for\,\,all}\,\,j.
\ee
The decoupling of $\CQ$-exact operators
is particularly important for the presence of topological observables in the theory. To see this explicitly, let us recall that the metric variation of the vacuum
expectation value (vev) of an operator $\CO$ is given by
\be
\label{dgCO}
\delta_g \left< \CO \right>= \int [\cD X] \left( \delta_g \CO -\CO\,\delta_g\mathcal{S}\right)\,e^{-\CS(X)}.
\ee
This variation vanishes if $\CO$ is
independent of the metric (or $\delta_g \CO$ is at least $\CQ$-exact), and if $\CO$
is $\CQ$-closed. With Equation (\ref{deltagL}), we arrive at the
fundamental and well-known statement that the Hilbert space $\CH$ of topological
observables is identified with the $\CQ$-cohomology,
\be
\CH=\mathrm{Ker}\,\CQ/\mathrm{Im}\,\CQ.
\ee
Such observables in Donaldson-Witten theory match the mathematically defined
Donaldson polynomials \cite{Witten:1988ze, DONALDSON1990257, Donaldson90}.

The validity of Equations (\ref{QV0}) and (\ref{dgCO}) requires a
careful analysis. Since the operator $\CQ$ can be expressed as a
derivative in field space  (see for example Equation
(\ref{Qdiff})), we
should anticipate that $\left<\{\CQ,\CO\} \right>$ might receive contributions from
boundaries or non-compact regions of field space. It turns out that some
simple choices of such $\CQ$-exact observables exhibit a worrisome divergent
contribution from such noncompact regions.  Indeed in the topologically twisted version of $\CN=2$
$SU(2)$ gauge theory, aka Donaldson-Witten theory \cite{Witten:1988ze},
many observables diverge near the singularities of the Coulomb
branch. Moreover, in asymptotically conformal
theories such as ${\rm SU}(2)$ gauge theory with $N_f=4$
\cite{Moore:1997pc}, and Argyres-Douglas theory
\cite{Moore:2017cmm}, boundary contributions are known to lead to a (continuous) metric
dependence of correlation functions.

We will analyze in this paper $\CQ$-exact observables on the Coulomb
branch of Donaldson-Witten theory. Even among $\CQ$-exact observables
which are regular on the interior of the $u$-plane, we will identify examples
whose correlation functions appear to diverge rather than
vanish. Motivated by this shortfall, we will put forward a new
prescription for the regularization and renormalization of correlation
functions. We will demonstrate that the prescription ensures the
decoupling of the $\CQ$-exact observables, while it is also consistent
with previous results \cite{Moore:1997pc, LoNeSha, Gottsche:1996aoa}.

To explain the new regularization, let us describe the contribution of the Coulomb branch to the path
integral in some more detail. The contribution of this branch is non-vanishing for four-manifolds with
$b_2^+\leq 1$, which provide a powerful arena for the analysis of
this phase of the theory. We will concentrate on four-manifolds with
$b_2^+=1$, for which the path integral reduces to an integral over the
order parameter
$u=\frac{1}{16\pi^2}\left<\mathrm{Tr}[\phi^2]\right>_{\mathbb{R}^4}$
\cite{Moore:1997pc, LoNeSha}, where
$\phi$ is the adjoint valued Higgs field of the theory, and
$\left<\dots\right>_{\mathbb{R}^4}$ denotes the vev in a normalized
vacuum state of the theory on $\mathbb{R}^4$.
The order parameter $u$ determines the effective coupling constant $\tau\in
\mathbb{H}$. Changing variables from $u$ to $\tau$ maps the $u$-plane
to six $\operatorname{SL}(2,\mathbb{Z})$ images of the fundamental domain $\CF_\infty=\mathbb{H}/\rm{SL}(2,\mathbb{Z})$
in the upper-half plane \cite{Moore:1997pc}. As a result, the path integral can be written as a sum of integrals of the form
\be
\label{intFmn}
L_{m,n,s}=\int_{\CF_\infty}d\tau\wedge d\bar \tau\, q^m \bar q^n y^{-s},
\ee
where $\tau = x + iy$ is the effective holomorphic coupling of the
theory. Such integrals have also appeared in the context of
one-loop amplitudes in string theory \cite{Lerche:1988np, Dixon:1990pc,
  Harvey:1995fq}, and much earlier in
mathematics as the (Petersson) inner product for cusp forms \cite{Petersson1950}.

The integral (\ref{intFmn}) is finite for $m+n>0$ and $s\in \mathbb{R}$, and
also for $m+n=0$ with $s>1$. The integrand however diverges exponentially for
$y=\rm{Im}(\tau)\to \infty$ if $m+n<0$. For a large class of such
$(m,n)$, namely when one of the two numbers is non-negative, the integral
can be evaluated using a, by now standard,
prescription \cite{Dixon:1990pc, Harvey:1995fq,
  Borcherds:1996uda}. Simply put, this prescription is to carry out first the integral over $x=\rm{Re}(\tau)$
and then the integral over $y$, such that
\be
\label{Imnreg}
L_{m,n,s}\sim \delta_{m,n} \int^\infty dy\, y^{-s} e^{-4\pi y n},
\ee
where we have just highlighted the potentially divergent part. The
$(m,n,s)$ encountered for the famous Donaldson-Witten observables in
the formulation of \cite{Moore:1997pc} are all such that this
regularization applies.

On the other hand, the condition that one element of the pair $(m,n)$ is non-negative,
may appear artificial, and as suggested above, we will present observables within Donaldson-Witten
theory which lead to integrals as (\ref{intFmn}) but with both $m$ and
$n<0$. The integrand in (\ref{Imnreg}) diverges in such cases, and the standard
prescription does not cure the infinity. The examples we present are
in fact $\CQ$-exact, such that the divergence leads to some tension with the expectation
that vacuum expectation values of  $\CQ$-exact operators vanish in topological
field theory. Rather than excluding these operators based on their
boundary behavior, we will demonstrate that they vanish once
appropriately regularized and renormalized. One observable we will study in this context is
\be
\label{tildeIintro}
\int_\bfx \{ \CQ, {\rm Tr}[\bar{\phi}\, \chi] \}=\int_\bfx \frac{d\bar u}{d\bar a}\, F_++\dots,
\ee
where $\bar \phi$ is the complex conjugate of the Higgs field $\phi$,  $\chi$ is
the self-dual Grassmann valued two-form field, $F_+$ is the self-dual
part of the curvature $F$ of the gauge connection and $\bfx$
is a two-cycle in the rational homology ring of $M$. The dots in (\ref{tildeIintro}) represent
terms involving fermions and the auxiliary field. This operator has
appeared
previously in the context of the CohFT
interpretation of Witten-like indices \cite{Moore:1998et}, and more recently
for the evaluation of Coulomb branch integrals using indefinite theta
functions in \cite{Korpas:2017qdo, Korpas:2018dag}.

This article proposes a new renormalization prescription for the $u$-plane
integral\footnote{With ``$u$-plane integral'', we refer to correlation
  functions on the Coulomb branch of rank one Donaldson-Witten theory, while ``Coulomb
 branch integral'' is used for arbitrary rank.},
which is based on the analytic continuation of the incomplete Gamma
function. This renormalization was recently developed by
Bringmann-Diamantis-Ehlen \cite{1603.03056} in
the context of modular integrals. See also \cite{bruinier2004} and
\cite{Duke2016}. For all $\CQ$-exact operators which are regular in
the interior of the $u$-plane, that is away from the strong and weak
coupling cusps, we show that this prescription
ensures the decoupling of $\CQ$-exact states from $\CQ$-closed
states. It reduces to the standard prescription described below
equation (\ref{intFmn}) where applicable, while it also could in principle be applied to evaluate correlation functions for
non-$\CQ$-closed observables. We hope that the new regularization makes the evaluation of new observables possible, and
that this will lead to further useful results concerning  topologically twisted theories and four-manifold topology.

The outline of this article is as follows. We give a brief
overview of Seiberg-Witten theory and its topologically twisted
formulation, Donaldson-Witten theory, in Section \ref{SWreview}. Section \ref{PathCor}
discusses the path integral and correlation functions of the
theory. Section \ref{regularization} introduces the
renormalization prescription,  which will be applied to the $\CQ$-exact observables on the $u$-plane in Section \ref{evaluation}. We
include various appendices with details on modular forms and some of the computations in the main
body of the paper.

%%%%%%%%%%%%%%%%%%%%%%%%%%%%%%%%%%%%%%%%%%%%%%
%%%%%%%%%%%%%%%%%%%%%%%%%%%%%%%%%%%%%%%%%%%%%%
%%%%%%%%%%%%%%%%%%%%%%%%%%%%%%%%%%%%%%%%%%%%%%
%%%%%%%%%%%%%%%%%%%%%%%%%%%%%%%%%%%%%%%%%%%%%%
%%%%%%%%%%%%%%%%%%%%%%%%%%%%%%%%%%%%%%%%%%%%%%
\section{Seiberg-Witten theory and Donaldson-Witten theory}
\label{SWreview}
This section gives a brief review of pure Seiberg-Witten theory
\cite{Seiberg:1994rs} with a rank one gauge group, and its topologically twisted counterpart known as
Donaldson-Witten theory \cite{Witten:1988ze}. We refer to \cite{Laba05, MooreNotes2017} for a detailed introduction to both.

\subsection{Seiberg-Witten theory}
Seiberg-Witten theory is the low energy effective theory of $\CN = 2$
supersymmetric Yang-Mills theory with gauge group ${\rm SU}(2)$ or ${\rm SO}(3)$. The theory contains a vector multiplet which consists of a
gauge field $A$, a pair of (chiral, anti-chiral) spinors $(\psi, \bar \psi)$, a complex scalar Higgs field
$\phi$ (valued in the complexification of the Lie algebra), and an auxiliary scalar field $D_{ij}$ (symmetric in $i$ and $j$) and possible matter representations. Here we
will consider pure Seiberg-Witten theory with gauge group as above,
which is broken to $\rm{U}(1)$ on the Coulomb branch $\CB$. The supersymmetry algebra of the theory contains a central charge $Z \in {\rm Hom}(\Gamma, \BC)$ where $\Gamma$ is the lattice of electric and magnetic charges of the theory fibered over $\CB$,
\[
Z(n_{\rm e},n_{\rm m})=n_{\rm e} a + n_{\rm m} a_D ,
\]
where $(n_{\rm e},n_{\rm m})\in \Gamma$ is the pair of electric-magnetic
charges, and the pair $(a,a_D)\in \mathbb{C}^2$ are the central charges for a
unit electric or magnetic charge. The central charge determines the
mass of BPS states, $m_{\text{BPS}}=|Z|$.

\begin{figure}[h]
\begin{center}
\includegraphics[width=16cm]{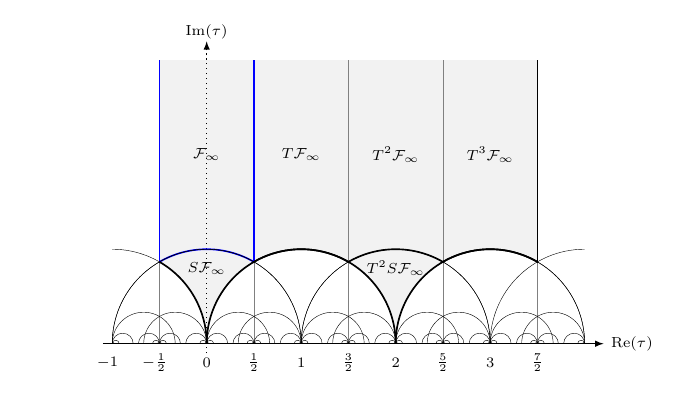}
\caption{Upper-half plane $\mathbb{H}$ with the area bounded by blue
   ($\CF_\infty$) a fundamental domain of $\mathbb{H}/{\rm
     SL}(2,\BZ)$, and the shaded area a fundamental domain of
   $\mathbb{H}/\Gamma^0(4)$. \label{fund_domain}}
\end{center}
\end{figure}

The Coulomb branch parameter $a$ and its dual $a_D$ are related by the holomorphic
prepotential $\CF$ of the theory
\be
a_D=\frac{\partial \CF(a)}{\partial a},
\ee
which in turn determines the effective coupling constant $\tau = \frac{\theta}{\pi} + \frac{8\pi i }{g^2} \in \BH$,
\be
\tau = \frac{\partial^2 \CF(a)}{\partial a^2}.
\ee
where $\theta$ is the instanton angle, $g$ the Yang-Mills coupling and
$\BH$ is the complex upper half-plane. The Coulomb branch $\CB$ is
parametrized by a single order parameter $u$,
\be
\label{uorder}
u = \frac{1}{16\pi^2}\left\bra \text{Tr}[\phi^2] \right\ket_{\mathbb{R}^4},
\ee
where the subscript indicates that this is a vev in a
normalized vacuum state of the theory on $\mathbb{R}^4$. The
renormalization group flow relates the Coulomb branch parameter $u$ and the effective coupling
constant $\tau$. Using the Seiberg-Witten geometry, the order parameter $u$ can be exactly expressed as function
of $\tau$ in terms of modular forms,
\be
\label{utau}
\frac{u(\tau)}{\Lambda^2} = \frac{\vartheta_2^4 + \vartheta_3^4}{2 \vartheta_2^2 \vartheta_3^2} = \frac{1}{8}q^{-\frac{1}{4}} + \frac{5}{2}q^{\frac{1}{4}} - \frac{31}{4}q^{\frac{3}{4}} + O(q^{\frac{5}{4}}),
\ee
where $\Lambda$ is a dynamically generated scale, $q=e^{2\pi i
  \tau}$, and the Jacobi theta functions $\vartheta_i(\tau)$ are explicitly given in Appendix
\ref{app_mod_forms}. The function $u(\tau)$ is invariant under
transformations $\tau \mapsto \frac{a\tau+b}{c\tau+d}$ for elements of the congruence subgroup $\Gamma^{0}(4) \subset
\mathrm{SL}(2,\BZ)$.\footnote{One way to understand this duality group
  is that the Seiberg-Witten curve of the theory
  is an elliptic curve for $\Gamma^0(4) \subset {\rm SL}(2,\BZ)$ \cite{Seiberg:1994rs}.}
 See Appendix \ref{app_mod_forms} for the definition of $\Gamma^0(4)$. This identifies the $u$-plane with a fundamental domain of
$\Gamma^0(4)$ in the upper-half plane $\mathbb{H}$, which we choose as
the images of the standard key-hole fundamental domain of $\mathrm{SL}(2,\BZ)$ under $\tau \mapsto \tau+1
$, $\tau + 2$, $\tau +3$, $\tau +4$, $-1/\tau$ and $2-1/\tau$. The
fundamental domains are displayed in Figure \ref{fund_domain}.

At the cusps $\tau\to 0$ (respectively $\tau\to 2$) a monopole (respectively a dyon) becomes massless, and the effective theory
breaks down since additional degrees of freedom need to be taken
into account. Another quantity which we will frequently encounter is
the derivative $\frac{da}{du}$, which can be expressed as function of $\tau$ as
\be
\label{dadu}
\frac{da}{du}(\tau) = \frac{\vartheta_2(\tau)\vartheta_3(\tau)}{\Lambda}.
\ee
It transforms under the generators of $\Gamma^0(4)$ as
\be
\label{dautrafos}
\begin{split}
&\frac{da}{du}(\tau+4)=-\frac{da}{du}(\tau),\\[0.4em]
&\frac{da}{du}\!\left( \frac{\tau}{\tau+1} \right)=(\tau+1)\, \frac{da}{du}(\tau).
\end{split}
\ee

\subsection{Donaldson-Witten theory}
\label{DonWitten}
Donaldson-Witten theory is the topologically twisted version of Seiberg-Witten theory with gauge group $\mathrm{SU}(2)$ or
$\rm{SO}(3)$ \cite{Seiberg:1994rs}. As mentioned in the introduction, topological twisting preserves a scalar fermionic nilpotent symmetry $\CQ$ of the $\CN=2$ Yang-Mills theory on an arbitrary four-manifold\footnote{Note that in \cite{Moore:1997pc, Korpas:2017qdo} this operator is denoted as $\overline{\CQ}$.}
\cite{Witten:1988ze}. For a four-manifold whose holonomy is
$\rm{SO}(4) \simeq \rm{SU(2)}_{-}\times \rm{SU}(2)_{+}$ the twist
replaces the initially flat ${\rm SU}(2)_R$ $R$-symmetry bundle, by the $\rm{SU}(2)_-$ subgroup of $\rm{SU}(2)_-\times
\rm{SU}(2)_+$, thereby changing the representations of operators under the
rotation group. The original supersymmetry generators transform as
the $(\bf{1,2,2})\oplus (2,1,2)$ representation of $\mathrm{SU}(2)_{+}
\times \mathrm{SU}(2)_{-} \times \mathrm{SU}(2)_R$ group. Their
representation under the twisted group $\mathrm{SU}(2)'_{+}
\times \mathrm{SU}(2)_{-} \times \rm{U}(1)_R$ is $({\bf{1,1}})^{+1} \oplus
({\bf{2,2}})^{-1} \oplus ({\bf{1,3}})^{+1} $. Each term in this direct
sum plays an important role in Donaldson-Witten theory.  The first
term $({\bf{1,1}})^{+1}$ corresponds to the BRST-type operator ${\CQ}$ we mentioned above, whose cohomology
provides topological invariants of the four-manifold. The second
term $({\bf{2,2}})^{-1}$ corresponds to the one-form operator
$K$. This operator provides a canonical solution to the descent equations \cite{Witten:1988ze}
\be
\label{CQO}
\{ \CQ, \CO^{(i+1)} \} = d \CO^{(i)}, \qquad i=0,\dots,3,
\ee
by setting $\CO^{(i)} = K^i \CO^{(0)}$  \cite{Labastida:1991qq, Moore:1997pc, LoNeSha}. Integration of these operators over
$i$-cycles gives topological observables since $\{ \CQ, K \} = d$.
  
Finally, we denote the third representation ${(\bf{ 1,3})}^{+1}$ by $L$,
which will be used to express the ${\CQ}$-exact operator
(\ref{tildeIintro}) in Section \ref{sectionI+}.\footnote{The operator $L$ is denoted 
  $Q_{\mu\nu}^+$ in Reference \cite[Section 2.1]{Ne}.} This operator
anti-commutes with the BRST supercharge to give $\{
\CQ,L\}=-(\bar{\sigma}_{mn})^{AB}\bar{Z}_{AB}\,dx^m \wedge dx^n$, where
$m,n$ are ${\rm SO}(4)$ indices while $A,B$ are ${\rm SU}(2)_R$
indices. We argue in Appendix \ref{Twisting} that for a compact K\"ahler surface $M$, this commutator can be written as
\be
\{ \CQ,L\}= \sqrt{2}i \bar{Z} J,
\ee
where $\bar{Z}:=\bar{Z}_{12}$ and $J \in \Omega^{1,1}(M)$ the K\"ahler
form associated with the metric $g$ of $M$.

The field content of the low energy topologically twisted theory is a one-form gauge
potential $A$, a complex scalar $a$, together with anti-commuting (Grassmann valued) self-dual
two-form $\chi$, one-form $\psi$ and zero-form $\eta$. The auxiliary
fields of the non-twisted theory combine to a self-dual two-form $D$. The action of the BRST operator $\CQ$ on these fields is given by
\begin{align} \label{barQcomm}   \nonumber
\{\mathcal{ Q},A\}&=\psi,           &  \{\mathcal{Q},a\} &= 0,              &  \{\mathcal{ Q},\bar a\} &=\sqrt{2}i \eta, \\[0.4em]
  \{\mathcal{ Q},D\}&=(d\psi)_+,         & \{\mathcal{Q},\eta\}&=0,  &  \{\mathcal{ Q},\psi\} &=4\sqrt{2}\,da,  \\[0.4em]  \nonumber
  \{\mathcal{ Q},\chi\} &=i(F_+-D).  & &         &   &
\end{align}
Later, it will be useful to express $\CQ$ as a derivative in field space,
\be
\label{Qdiff}
\CQ=\psi\,\frac{\partial }{\partial
  A}+(d\psi)_+\frac{\partial}{\partial D}+4\sqrt{2}da\,\frac{\partial}{\partial \psi}+\sqrt{2}i\eta\,\frac{\partial}{\partial \bar a} +i (F_+-D)\,\frac{\partial}{\partial \chi}.
\ee
The low energy Lagrangian of the Donaldson-Witten theory is given by \cite{Moore:1997pc}
\be
\label{Lagr}
\begin{split}
\mathcal{L}& =\frac{i}{16 \pi} (\bar \tau F_+ \wedge F_++\tau F_-\wedge F_-)+\frac{y}{8\pi} da\wedge * d\bar a-\frac{y}{8\pi} D\wedge * D\\
&\quad -\frac{1}{16\pi} \tau \psi \wedge * d\eta+\frac{1}{16\pi}\bar \tau \eta \wedge d*\psi+\frac{1}{8\pi}\tau \psi \wedge d\chi-\frac{1}{8\pi}\bar \tau \chi \wedge d\psi\\
& \quad +\frac{\sqrt{2}i}{16\pi} \frac{d\bar \tau}{d\bar a} \eta \chi \wedge (F_++D)-\frac{\sqrt{2}i}{2^7\pi}\frac{d\tau}{da}\psi\wedge \psi\wedge (F_-+D)\\
& \quad + \frac{i}{3\cdot 2^{11}} \frac{d^2\tau}{da^2} \psi\wedge \psi\wedge\psi\wedge\psi-\frac{\sqrt{2}i}{3\cdot 2^5 \pi}\left\{\CQ, \chi_{\mu\nu} \chi^{\nu\lambda} \chi_{\lambda}^{\,\,\mu} \right\}\sqrt{g}\,d^4x.
\end{split}
\ee

%%%%%%%%%%%%%%%%%%%%%%%%%%%%%%%%%%%%%%%%%%%%%%
%%%%%%%%%%%%%%%%%%%%%%%%%%%%%%%%%%%%%%%%%%%%%%
%%%%%%%%%%%%%%%%%%%%%%%%%%%%%%%%%%%%%%%%%%%%%%
%%%%%%%%%%%%%%%%%%%%%%%%%%%%%%%%%%%%%%%%%%%%%%
%%%%%%%%%%%%%%%%%%%%%%%%%%%%%%%%%%%%%%%%%%%%%%

%%%%%%%%%%%%%%%%%%%%%%%%%%%%%%%%%%%%%%%%%%%%%%
%%%%%%%%%%%%%%%%%%%%%%%%%%%%%%%%%%%%%%%%%%%%%%
%%%%%%%%%%%%%%%%%%%%%%%%%%%%%%%%%%%%%%%%%%%%%%
%%%%%%%%%%%%%%%%%%%%%%%%%%%%%%%%%%%%%%%%%%%%%%
%%%%%%%%%%%%%%%%%%%%%%%%%%%%%%%%%%%%%%%%%%%%%%
\section{Correlation functions of $\CQ$-exact observables} \label{PathCor}
We start in this section an analysis of correlation functions of
$\CQ$-exact observables to verify the Ward identities (\ref{QV0}) and
(\ref{CQdecoupling}). After collecting a few useful facts about
four-manifolds with $b_2^+=1$, we recall the contribution to the
path integral of the $u$-plane in Subsection \ref{path_int_Don}. In the remainder of the section,
we discuss $\CQ$-exact observables at an increasing level of
generality. We summarize our findings in Subsection \ref{CorrSum}.

\subsection{Four-manifolds with $b_2^+=1$}
Let $M$ be a smooth, simply connected, compact four-manifold without
boundary. Its basic topological numbers are its Euler character
$\chi(M)= 2-2b_1(M)+b_2(M)$ and signature
$\sigma(M)=b_2^+(M)-b_2^-(M)$, where $b_1(M) ={\rm
  dim}(H_1(M,\BR)) $ and $b_2^{\pm}(M) = {\rm
  dim}(H_2(M,\BR)^{\pm})$. We will omit the dependence on $M$ unless a
confusion may arise. We will restrict in the following to four-manifolds with
$b_2^+=1$, since the $u$-plane integral only contributes for this
class of four-manifolds. A four-manifold $M$ with $b_2^+=1$ admits an almost complex
structure, since any simply connected
four-manifold with $b_2^{+}$ odd does \cite{Donaldson90}. We denote the canonical class of $M$ by $K_M \in H^2(M,\mathbb{Z})$, which equals
the second Stiefel-Whitney class modulo $H^2(M,2\BZ)$.

The intersection form on the middle cohomology  provides a
natural bilinear form $B : H^2(M,\mathbb{R}) \times H^2(M,\mathbb{R})
\to \mathbb{R}$ that pairs degree two co-cycles,
\be
B(\bfk_1, \bfk_2) := \int_M \bfk_1\wedge \bfk_2,
\ee
and whose restriction to $H^2(M,\mathbb{Z}) \times H^2(M,\mathbb{Z})$
is an integral bilinear form with signature $(1,b_2-1)$. The bilinear form provides the quadratic
form $Q(\bfk) :=  B(\bfk, \bfk)\equiv \bfk^2$, which can be brought to
a simple standard form \cite[Section 1.1.3]{Donaldson90}. We denote
the period point by $J$, i.e. the harmonic two-form, satisfying
\be
* J=J \in H^2(M,\mathbb{R}),\qquad  J^2=1,
\ee
with $*$ the Hodge $*$-operation.  Using the period point, we can decompose elements $\bfk\in H^2(M)$ to
its self-dual and anti-self-dual components:
$\bfk_+=B(\bfk,J)\,J$ and $\bfk_-=\bfk-\bfk_+$, the anti-self-dual part of $\bfk$.
For later use, we mention that the canonical class is a characteristic vector of $H^2(M,\mathbb{Z})$ and satisfies
\be
K_M^2=\sigma+8.
\ee

\subsection{Path integral of the Donaldson-Witten theory}
\label{path_int_Don}

We consider Donaldson-Witten theory on a four-manifold $M$ with
$b_2^+=1$ as detailed above. We choose a
fixed 't Hooft flux $[\mathrm{Tr} (F/4\pi)]=2\bfmu\in H^2(M,\mathbb{Z})$ for the gauge bundle (we think of $H^2(M,\mathbb{Z})$ as a lattice in $H^2(M,\mathbb{R})$, thus modding out by torsion). 
Then we can divide by $2$ and in particular $\bfmu$ can be half-integral.) 
The Coulomb branch integral $\Phi_{\bfmu}^J$ \cite{Moore:1997pc} of Donaldson-Witten
theory, without any operator insertions, is defined as the usual path integral over the infinite
dimensional field space,
\bes
\Phi_{\bfmu}^J = \int [\cD a\, \cD \bar a\, \cD A\, \cD \eta\, \cD \psi\, \cD \chi\,  \cD D ]\, e^{-\CS}  \equiv \bra 1 \ket.
\ees

We will review in this subsection that the path integral is
well-defined and reduces to a modular integral over the domain
$\mathbb{H}/\Gamma^0(4)$. For the chosen class of four-manifolds, $\Phi_{\bfmu}^J$ reduces to a finite dimensional integral over
the zero modes \cite{Moore:1997pc}. For simplicity,
we restrict to simply connected manifolds, $\pi_1(M) = 0 $ and
therefore $b_1(M)=0$, which do not admit $\psi$ zero modes. The
path integral of the effective theory on the Coulomb branch becomes then
\be \label{PartitionFunction}
\Phi_{\bfmu}^J=\sum_{{\rm U}(1)\,\,{\rm fluxes } } \int da\wedge
d{\bar{a}}\wedge d\eta\wedge d\chi \wedge dD\, \,A(u)^{\chi(M)}\,B(u)^{\sigma(M)}\,e^{-\int_M\CL_0},
\ee
where the $a$, $\bar a$ and $\eta$ denote now the zero-modes,
i.e. they are constant functions on $M$. The Lagrangian
$\mathcal{L}_0$ is $\CL$ (\ref{Lagr}) restricted to
the zero modes including the ones of the gauge field.

The functions $A(u)$ and $B(u)$ are curvature
couplings; they are holomorphic functions of $u$, given by \cite{Moore:1997pc, Witten:1995gf}
\be
\begin{split}
A(u) &= \alpha \left( \frac{du}{da} \right)^{\frac{1}{2}}, \\[0.5em]
B(u) &= \beta\, (u^2 - 1)^{\frac{1}{8}}, \
\end{split}
\ee
where we have set $\Lambda=1$, and $\alpha$ and $\beta$ are numerical
factors. In more general theories including matter, such as the $N_f=4$ theory, they may depend on
parameters such as masses and coupling constants.

To integrate over the auxiliary field, let us introduce the Lagrangian
$\CL_{0,D}$, which consists of the terms in $\CL_0$ involving $D$,
\be
 \mathcal{L}_{0,D}=-\frac{y}{8\pi}  D\wedge D+\frac{\sqrt{2}i}{16\pi} \frac{d\bar \tau}{d\bar a}\eta\chi\wedge D .
\ee
After a Wick rotation $D\to iD$, the Gaussian integration over its zero mode yields
\be
\label{GaussDint}
\int dD\, e^{-\int_M \CL_{0,D}}=2\pi i
\sqrt{\frac{2}{y}}.
\ee
The only remaining term involving fermion zero-modes in $\CL_0$ is
\be \label{FermionicLagrangian}
\mathcal{L}_{0,f }=\frac{\sqrt{2}i}{16\pi} \frac{d\bar \tau}{d\bar a} \eta
\chi\wedge F_+,
\ee
such that integrating over the $\eta$ and $\chi$ zero modes gives
\be
\label{intfermionzeros}
\int d\eta \wedge d\chi ~e^{-\int_M \mathcal{L}_{0,f}}=\frac{\sqrt{2}i}{4}
\frac{d\bar \tau}{d\bar a} B(\bfk,J),
\ee
where the vector $\bfk$ equals the $\rm{U}(1)$ flux $[F]/4\pi\in H^2(M,\BZ)+\bfmu$.

The sum over $\bfk$ in (\ref{PartitionFunction}) takes the
form of a Siegel-Narain theta function
\be \label{PsiK}
\Psi^J_\bfmu\left[ \CK_{\rm p} \right](\tau,\bar \tau)=\sum_{\bfk\in \Lambda +
  \bfmu}\,\CK_{\rm p}(\bfk)\, (-1)^{B(\bfk,K_M)}\,q^{-\frac{\bfk_-^2}{2}} \bar q^{\frac{\bfk_+^2}{2}},
\ee
where the kernel $\CK_{\rm p}$ equals
\be
\label{CK0}
\CK_{\rm p}(\bfk)=-\frac{\pi}{\sqrt{y}}\,B(\bfk, {J}),
\ee
which follows from multiplying
(\ref{GaussDint}) and (\ref{intfermionzeros}), and dividing by the
factor $\frac{d\bar \tau}{d\bar a}$ since this
provides the change of variables from the Coulomb branch parameter $a$
to $\tau\in \mathbb{H}/\Gamma^0(4)$. When we consider correlation
functions in the next subsection, we will find different expressions
for the kernel depending on the inserted fields. Appendix
\ref{SNtheta} lists a number of useful properties of
$\Psi^J_\bfmu[1]$.

We can express the integrand in (\ref{PartitionFunction}) more compactly, using Matone's
formula \cite{Matone:1995rx}
\be
\frac{du}{d\tau} = \frac{4\pi }{i}(u^2-1)\left( \frac{da}{du} \right)^2,
\ee
and the identities (\ref{utau}) and (\ref{dadu}). This gives for $\Phi_\bfmu^J$
\be \label{PartFun}
\Phi_\bfmu^J=\int_{\BH/\Gamma^0(4) } d\tau \wedge d\bar{\tau} \,
\tilde{\nu}(\tau)\, \Psi_{\bfmu}^J[\CK_{\rm p}](\tau, \bar{\tau}),
\ee
with
\be  \label{measureterm}
\tilde{\nu}(\tau) := -8i (u^2-1)\frac{da}{du}\vartheta_4(\tau)^{\sigma},
\ee
and whose modular transformations for the two generators $ST^{-1}S: \tau
\mapsto \frac{\tau}{\tau+1}$ and $T^4: \tau \mapsto \tau+4$ of
$\Gamma^0(4)$ are:
\be
\begin{split}
&\tilde {\nu}\!\left(\frac{\tau}{\tau+1}\right)=(\tau+1)^{2-b_2/2} e^{-\frac{\pi i \sigma}{4}} \tilde {\nu}(\tau),\\[0.5em]
&\tilde {\nu}(\tau+4)=-\tilde {\nu}(\tau).
\end{split}
\ee
The measure $\tilde{\nu}(\tau)$ behaves near the weak coupling cusp
$\tau\to i\infty$ as $\sim q^{-\frac{3}{8}}$, and near the monopole
cusp, $\tau_M=-1/\tau\to i\infty$ as $\sim q_M^{1+\frac{\sigma}{8}}$.

An important requirement for (\ref{PartFun}) is modular
invariance of the integrand under $\Gamma^0(4)$ transformations. We can easily
determine the modular transformations of $\Psi_{\bfmu}^J[\CK_{\rm p}]$ from
those of $\Psi_{\bfmu}^J[1]$ (\ref{Psitrafos}). The effect of
inserting the kernel $\CK_{\rm p}$ in $\Psi_{\bfmu}^J[1]$  is to increase
the weight by $(\frac{1}{2},\frac{3}{2})$. (The factor $1/\sqrt{y}$
contributes $(\frac{1}{2},\frac{1}{2})$ and $B(\bfk,J)$
contributes $(0,1)$ to the total weight.) We then arrive at
\be
\begin{split}
&\Psi_{\bfmu}^J[\CK_{\rm p}]\!\left(\frac{\tau}{\tau+1},\frac{\bar
    \tau}{\bar \tau+1} \right) = (\tau+1)^{\frac{b_2}{2}}
(\bar{\tau}+1)^2 e^{\frac{\pi i}{4}\sigma}\, \Psi_{\bfmu}^J[\CK_{\rm p}](\tau,\tau),\\[0.5em]
&\Psi_{\bfmu}^J[\CK_{\rm p}](\tau+4,\bar \tau+4) =e^{2\pi i
  B(\bfmu,K)}\, \Psi_{\bfmu}^J[\CK_{\rm p}](\tau,\bar \tau),
\end{split}
\ee
where we used that $K_M^2=8+\sigma$. We see that the integrand of
(\ref{PartFun}) is invariant under the $\tau\mapsto
\frac{\tau}{\tau+1}$ transformation. However, if $B(\bfmu, K)=0 \mod
\mathbb{Z}$, the $\tau\mapsto \tau+4$ does multiply the integrand by
$-1$, but one can show that $\Psi_{\bfmu}^J[\CK_{\rm p}]$ vanishes in this
case, such that there is no violation of the duality. We conclude therefore that the Coulomb branch
integral (\ref{PartFun}) is well-defined since the measure $d\tau
\wedge d\bar{\tau}$ transforms as a mixed modular form of weight $(-2,-2)$ while
the product $\tilde{\nu}\, \Psi_{\bfmu}^J[\CK_{\rm p}]$ is a mixed modular
form of weight (2,2) for the congruence subgroup $\Gamma^0(4)$.

The evaluation of $\Phi^J_\bfmu$ will be discussed in more detail in
upcoming work \cite{to_appear}. We will continue in the next subsection by considering
correlation functions of BRST exact observables, which need to
satisfy the same requirements of modular invariance of the
integrand as above. We summarize in Table \ref{Table} the weights
of the various ingredients that appear in $u$-plane
integral for future use.

\begin{center}
\begin{table}[]
\centering
\begin{tabular}{|l@{\hspace{0.4in}} | r|}  \hline
Ingredient & Mixed weight   \\ \hline
$d \tau \wedge d\bar{\tau}$ & $(-2,-2)$  \\
 $y$ & $(-1, -1)$ \\
  $\partial_{\bar{\tau}}\,f$ & $(k,2)$ if $f$ has weight $(k,0)$ \\
 $\tilde{\nu}(\tau)$ & $ (2-b_2/2,0) $ \\
 $ \Psi_{\bfmu}^{J}[1] $ &  $\frac{1}{2}((b_2-1),1)$   \\ \hline
\end{tabular}
\caption{Various modular weights for the $u$-plane
  integral. Transformations are in $\mathrm{SL}(2,\mathbb{Z})$ for the first three rows, while in
  $\Gamma^0(4)$ for the last two rows. \label{Table}}
\end{table}
\end{center}

\subsection{An anti-holomorphic $\CQ$-exact observable}
\label{sectionI+}
We will analyse in this subsection the $u$-plane integral with the
insertion of a specific anti-holomorphic $\CQ$-exact
surface observable. Our analysis will demonstrate that its vev appears
to diverge rather than vanish as suggested by the Ward-Takahashi
identity, which will motivate the new regularization in the next section.

The observable of interest is
\be
\begin{split}
I_{+ }(\bfx)&=-\frac{1}{4\pi}\int_\bfx  \left\{\CQ, \left\{ L,
    \mathrm{Tr}[\bar \phi^2] \right\} \right\} \\
             &= -\frac{1}{4\pi}\int_\bfx  \left\{\CQ, {\rm Tr}[ \bar{\phi}\,\chi]  \right\}, \
\end{split}
\ee
where $\bfx \in H_2(M,\BQ)$ is a two-cycle, and $L$ is the twisted supersymmetry generator
discussed in Section \ref{DonWitten}. The subscript $+$ is to indicate that it
involves a self-dual two-form field, and is in a sense a self-dual
counterpart of the holomorphic, anti-self dual Donaldson observable $I_-(\bfx)$
\cite{Korpas:2017qdo}. Using the action of $L$, we can determine the image of $I_{+
}(\bfx)$ in the IR theory, denoted by $\widetilde I_{+ }(\bfx)$, in terms of the IR fields,
\be
\label{tildeI}
\begin{split}
\widetilde I_{+}(\bfx)&=-\frac{1}{4\pi} \int_{\bfx} \left\{ \mathcal{
    Q},\frac{d\bar u}{d\bar a} \chi \right\} \\
&=-\frac{i}{\sqrt{2}\pi} \int_{\bfx}\left(  \frac{1}{2} \frac{d^2\bar
  u}{d\bar a^2}\eta\,\chi+\frac{\sqrt{2}}{4} \frac{d\bar u}{d\bar
  a}(F_+-D) \right).
\end{split}
\ee

As for the partition function, we first integrate over the $D$ zero
mode using (\ref{GaussDint}) and
\be
\label{Dint2}
\int dD\,\left[\int_\bfx D\right] e^{-\int_M \mathcal{L}_{0,D}}=2\pi i \sqrt{\frac{2}{y}}\,\left(\frac{\sqrt{2}i}{4y}\frac{d\bar \tau}{d\bar a}\,\int_\bfx \eta \chi\right).
\ee
Next, we integrate over the fermion zero modes; $\int d\eta\wedge
d\chi \wedge dD\,
\widetilde I_+(\bfx)\, e^{-\int_M \mathcal{L}_{0,f}+\CL_{0,D}}$ evaluates to
$\frac{d\bar{\tau}}{d\bar{a}}\, \CK_{+}(\bfk)$, with
$\CK_{+}(\bfk)$ given by
\be \label{Ker2}
\CK_{+}(\bfk) := \frac{2\,B(\bfx,J)}{\sqrt{y}}  \left( -\frac{1}{2} \frac{d^2\bar u}{d\bar a\, d \bar \tau} +\frac{i}{8\, y}
  \frac{d\bar u}{d\bar a}  +\frac{\pi
  i}{2} \frac{d\bar u}{d\bar a}\, \bfk_+^2 \right).
\ee
Once combined with the sum over the $\rm{U}(1)$ fluxes, Equation
(\ref{tildeI}) can be written in a compact form. One arrives at a total
derivative with respect to $\bar \tau$,
\be \label{Ker33}
\Psi_{\bfmu}^{J}[\CK_{+}](\tau, \bar{\tau}) = -\partial_{\bar \tau} \left( \frac{B(\bfx,J)}{\sqrt{y}}\,  \frac{d\bar{u}}{d\bar{a}}\,\Psi_{\bfmu}^{J}[1](\tau,\TB) \right).
\ee
This expression demonstrates that $\Psi_{\bfmu}^{J}[\CK_{+}]$ vanishes for
$B(\bfmu,K_M)=\frac{1}{2} \mod \BZ$, since $\Psi_{\bfmu}^{J}[1]$
vanishes in this case. If non-vanishing
$\Psi_{\bfmu}^{J}[\CK_{+}]$ has the required modular properties: it
transforms with modular weight $(\frac{b_2}{2}, 2)$, and changes by a sign under $\tau\mapsto \tau+4$.

For the one-point function of $\widetilde I_+(\bfx)$, we arrive at the integral
\be
\langle \widetilde I_+(\bfx)\rangle=-\int_{\BH/\Gamma^{0}(4)} d\tau\wedge d\bar
\tau \,\partial_{\bar \tau}\left( \widetilde \nu\, \frac{B(\bfx,J)}{\sqrt{y}}\,  \frac{d\bar{u}}{d\bar{a}}\,\Psi_\bfmu^J[1] \right).
\ee
We can easily evaluate this integral using Stokes' theorem. This reduces to arcs close to the three
cusps of $\mathbb{H}/\Gamma^0(4)$, $\tau\to i\infty$, $0$ and $2$. But here
is where the surprise occurs: since $\frac{d\bar u}{d\bar a}$ diverges as
$\bar q^{-\frac{1}{8}}$ for $\tau\to \infty$ and $\tilde \nu(\tau)$ as
$q^{-\frac{3}{8}}$, we find integrals $L_{m,n,s}$ (\ref{intFmn}) with both $m$
and $n<0$\, for the cusp at $i\infty$! The standard prescription mentioned above
Equation (\ref{Imnreg}) therefore does not cure the divergence if
$\Psi_\bfmu^J\sim q^{\frac{1}{4}}$ for $\tau\to i\infty$. We will explain in Section \ref{evaluation} that the integral can be
 properly renormalized. This will result in $\bra \widetilde{I}_{+}(\bfx)
 \ket=0$, in agreement with the global BRST symmetry.

\subsection{General $\CQ$-exact observables}
\label{GeneralQexact}
Motivated by the example $\widetilde I_+$, we will make an analysis of
general $\CQ$-exact observables in this subsection. We assume that these observables satisfy the constraints of
single-valuedness on the $u$-plane, and that these would be automatically
satisfied if we derive them from $\CQ$-exact UV operators. We
will find that the $u$-plane integrands for such observables always take the form of a
total $\bar \tau$-derivative, which facilitates their evaluation in
Section \ref{evaluation}.

We grade the $\CQ$-closed and exact observables by their form degree:

\subsubsection*{0-form operators}
A general 0-form operator $\CO_0$ can be written as
\be
\label{Q0form}
\CO_0=V_0(a,\bar a)+V_1(a,\bar a)\,\eta,
\ee
where we require that $V_j$ are real-analytic functions in the real and imaginary part
of $a$ on the interior of the $u$-plane, in other words, the $V_j$ do not have singularities away from the weak and strong coupling
cusps in the $u$-plane. After acting with $\CQ$ (\ref{barQcomm}) on this expression, we find that the most general $\CQ$-exact 0-form operator is
\be
\label{Q0exact}
\begin{split}
G_0& =\{\CQ,\CO\}= \sqrt{2}i\, \partial_{\bar a}  V_0(a,\bar a)\,\eta.
\end{split}
\ee
The vev $\langle G_0 \rangle$ vanishes after integration over the fermionic modes, since $G_0$ is
Grassmann odd and the action only contains Grassmann even
terms. We thus find that any 0-form operator satisfies the Ward
identity (\ref{QV0}). Moreover, any product $\prod_j \CO_{0,j}$ with
$\CO_{0,j}$ $\CQ$-exact 0-form operators is also of the form (\ref{Q0form}).

Let us next consider $\CQ$-closed 0-form observables.
We deduce from (\ref{Q0exact}) that for any $\CQ$-closed
observable the $\eta^0$ term is necessarily holomorphic, thus
\be
\label{0closed}
C_0=W_0(a)+W_1(a,\bar a)\,\eta,
\ee
where $W_j$ are again real-analytic functions on the interior of the
$u$-plane. For single
valuedness of the $u$-plane integrand, $W_0$ must be invariant under
$\Gamma^0(4)$ transformations. For example the famous point operator
is $u$. Comparing (\ref{Q0exact}) and (\ref{0closed}), we deduce
that there exist  $\CQ$-closed forms, linear in $\eta$, which are not
$\CQ$-exact. They do however not contribute to correlation functions
since they are Grassmann odd. For the same reason, the Ward-Takahashi identity (\ref{CQdecoupling}) is satisfied for $0$-form
observables: for any 0-form observable $\CO_0$,
\be
\langle \{\CQ,\CO_0\} \prod_{j} \CO_{0,j}\rangle=0,
\ee
if all $\CO_{0,j}$ are $\CQ$-closed 0-form observables.

 \subsubsection*{2-form operators}
We continue with $\CQ$-exact 2-form operators $G_2=\{\CQ,\CO_2\}$. We
let $\CO_2$ be the most general 2-form field, expressed as
\be
\CO=\sum_{X\,\in \{\chi,\, F_\pm,\,D,\,\psi\wedge \psi\}} V_{X,j}(a,\bar a)\,\eta^j\,X.
\ee
where $V_{X,j}(a,\bar a)$ are again real-analytic functions without
singularities away from the strong and weak coupling singularities. Comparing with Equation (\ref{tildeI}),
we find that for $\widetilde I_+$ the function $V_{\chi,0}(a,\bar a)$ that
\be
V_{\chi,0}(a,\bar a) = -\frac{1}{4\pi}\frac{d\bar u}{d\bar a},
\ee
and all other $V_{X,j}$ equal to 0. Acting with $\CQ$ on $\CO$ gives the following expression
for $G_2$
\be
\begin{split}\label{MostGeneralQ}
G_2&=\sqrt{2}i\,\partial_{\bar a} V_{\chi,0} \eta\,\chi + i\,
V_{\chi,0} \,(F_+-D)\\
&\quad +\sqrt{2}i\,\partial_{\bar a} V_{F_\pm,0} \eta\,F_\pm +
V_{F_\pm,0} \,(d\psi)_\pm\\
&\quad + \sqrt{2}i\,\partial_{\bar a} V_{D,0} \eta\,D +  V_{D,0} \,(d\psi)_+\\
&\quad + \sqrt{2}i\,\partial_{\bar a} V_{\psi\wedge\psi,0}
\eta\,\psi\wedge\psi -  8\sqrt{2}\,V_{\psi\wedge\psi,0}\, \psi\wedge da\\
&\quad -iV_{\chi,1}\eta\,\,(F_+-D)-V_{F_\pm,1}\eta\,(d\psi)_\pm\\
&\quad - V_{D,1}\eta\,(d\psi)_+-4\sqrt{2}\,V_{\psi\wedge \psi, 1}\eta\,\psi\wedge da,
\end{split}
\ee
where in the second and fifth line $\pm$ represents a sum over $+$ and $-$.

In correlation functions, we integrate $G_2$ over a two-cycle $\bfx\in
H_2(M,\mathbb{Z})$. For simplicity of notation, we set
\be
G_2(\bfx)\equiv\int_{\bfx} G_2.
\ee
To evaluate $\langle G_2(\bfx) \rangle$ for the class of
four-manifolds relevant to this paper, we reduce to zero modes and
integrate over the $\eta$ and $\chi$ zero-modes. This ensures that all all terms on the right hand side of
(\ref{MostGeneralQ}) have a vanishing contribution to $\langle G_2(\bfx)
\rangle$, {\it except} the two terms with
$V_{\chi,0}$. We will proceed with only these two terms, which is
similar to the analysis in Section \ref{sectionI+}. Integrating over
$D$ gives
\be
\begin{split}\label{MostGeneralQ}
&\int dD\,G_2(\bfx)\,e^{- \int_M \mathcal{L}_{0,D}}= \\
&\qquad 2\pi i \sqrt{\frac{2}{y}}\left[ \int_\bfx \left(\sqrt{2}i\,\partial_{\bar a} V_{\chi,0} +  \frac{\sqrt{2}}{4y}\frac{d\bar \tau}{d\bar a}V_{\chi, 0}  \right) \eta \chi + 4\pi i V_{\chi,0}\, B(\bfk_+,\bfx)\right].
\end{split}
\ee
Integrating subsequently over the $\eta$ and $\chi$ zero modes gives
the sum over fluxes $\Psi_\bfmu^J[\CK_2]$, with kernel
\be
\label{CK2}
\CK_{2}\ =-  \frac{4 \pi}{\sqrt{y}}\, B(\bfx,J)   \left(  \partial_{\bar
    \tau}V_{\chi,0} - \frac{i}{4y} V_{\chi,0}  - \pi i  V_{\chi,0}\bfk_+^2  \right),
\ee
which can be simplified to
\be
\Psi_{\bfmu}^J[\CK_{2}](\tau, \bar \tau) =\partial_{\bar \tau}\left( \frac{ 4\pi B(\bfx,J)}{\sqrt{y}}V_{\chi,0}
\Psi_{\bfmu}^J[1](\tau,\bar \tau) \right).
\ee
We can easily deduce the modular properties of $V_{\chi,0}$ necessary
for single-valuedness of the integrand: $V_{\chi,0}$ must have weight
$(0,-1)$, and transform with the same multiplier system as $d \bar u/d\bar a$.

Our next aim is to consider correlation functions of a $\CQ$-exact
operator with a $\CQ$-closed operator. To this end, let us analyze the form of the most general $\CQ$-closed  two-form
operator $C_2=\sum_X W_{X,j} X\,\eta^j$. Since we
integrate over a closed two-cycle, $\int_\bfx \CO$, the right hand side of
(\ref{MostGeneralQ}) can vanish up to a total derivative. The relations
this imposes on the functions $W_{*,j}$ are easily read off from
(\ref{MostGeneralQ}). We obtain
\begin{align} \nonumber
&W_{\psi\wedge\psi,0}=\frac{1}{8\sqrt{2}}\partial_a W_{F_-,0},     \qquad  W_{\chi,1} =\sqrt{2}\partial_{\bar a} W_{F_+,0},  &   &              \\
&W_{F_-,0} =W_{F_+,0}+W_{D,0}         = \text{holomorphic}, &&   \\[0.4em] \nonumber
&W_{D,1} + W_{F_+,1}=W_{F_-,1}  =W_{\psi\wedge \psi,1} = W_{\chi,0} =0. &   &
\end{align}
Then we see that $C_2$ can be expressed as
\be
\label{COclosed}
\begin{split}
C_2&=W_{F_-,0}\,(F_-+D)+\frac{1}{8\sqrt{2}}\partial_a
W_{F_-,0}\,\psi\wedge \psi\\
&\quad +\sqrt{2}\partial_{\bar a}W_{F_+,0} \eta \chi +W_{F_+,0}\,(F_+-D) \\
&\quad +W_{F_+,1}(a,\bar a)\,\eta\,(F_+-D),
\end{split}
\ee
and
\be
\{\CQ,C_2\}=d\left( W_{F_-,0}(a)\,  \psi \right).
\ee
The first two terms in (\ref{COclosed}) are holomorphic and do match the terms in the standard
Donaldson surface observable as derived using the descent
formalism. Comparing with Equation (2.17) in \cite{Moore:1997pc}, we
find that for the surface observable\footnote{Note that we find
  a different sign of the $\psi\wedge \psi$ term compared to
  \cite{Moore:1997pc, Korpas:2017qdo}.}
\be
W_{F_-,0}(a) = -\frac{i}{4\pi} \frac{du}{da}.
\ee
The last three terms on the rhs of (\ref{COclosed}) are $\CQ$-exact,
and the first two are of form which do not automatically vanish.

We consider next the correlation function for the product of a
$\CQ$-exact and a $\CQ$-closed operator, $G_2(\bfx)\,
C_2(\bfx')$. As before, the path integral restricts to zero modes and
Grassmann even terms. Integration over the zero modes of $D$, $\eta$
and $\chi$ gives
\be
\label{G2C2}
\begin{split}
&\int d\eta \wedge d\chi\wedge dD\,G_2(\bfx)\, C_2(\bfx')\, e^{-\int_M \CL_{0,D}+\CL_{0,f}} =\\
&\quad  16\pi^2\, \frac{d\bar \tau}{d\bar a}\, B(\bfx,J)\,B(\bfx',J)\,B(\bfk,J)\left[ \left( \partial_{\bar \tau} \frac{V_{\chi,0}W_{F_+,0}}{\sqrt{y}} \right) - \frac{ V_{\chi,0}W_{F_+,0}}{\sqrt{y}}\,  \pi i\bfk_+^2\right]\\
&\quad  +16\pi^2\, \frac{d\bar \tau}{d\bar a}\, W_{F_-}\,B(\bfx,J)\,B(\bfx',\bfk_-)\,\left[ \left( \partial_{\bar \tau} \frac{V_{\chi,0}}{\sqrt{y}} \right) - \frac{ V_{\chi,0}}{\sqrt{y}}\,  \pi i\bfk_+^2\right],
\end{split}
\ee
where the first line is due to the product of the non-holomorphic $\CQ$-exact part of
$C_2$ with $G_2$, and the second line is the contribution of the
product from the
holomorpic part of $C_2$ with $G_2$. Note that the term in brackets on
the second line is very similar to (\ref{CK2}), since the holomorphic part
commutes with $\partial_{\bar \tau}$. Using (\ref{G2C2}), we may write
the $u$-plane integrand for $\langle
G_2(\bfx)\,C_2(\bfx')\rangle$ as\footnote{We have divided by $i$ in the first kernel of $\Psi_\bfmu^J$ for consistency with Equation (\ref{Kell}) at the end of this subsection.}
\be
\begin{split}
&B(\bfx,J)\, B(\bfx',J)\, \partial_{\bar \tau}\! \left(\widetilde \nu \, V_{\chi,0}\,W_{F_+,0}\,
  \Psi_\bfmu^J[\CK_2^{(2)}/i] \right)\\
&+B(\bfx,J)\,\partial_{\bar \tau}\!\left( \widetilde \nu\,W_{F_-,0}\,
  V_{\chi,0}\,
  \Psi_\bfmu^J[\CK_-] \right),
\end{split}
\ee
where the kernels read
\be
\begin{split}
\CK_{2}^{(2)} &=\frac{16\pi^2 i}{\sqrt{y}}\, B(\bfk,J),\qquad
\CK_-=\frac{16\pi^2}{\sqrt{y}} B(\bfx',\bfk_-).
\end{split}
\ee
We have thus demonstrated that the $u$-plane integral takes for this
correlator also the form of a total derivative. We will not develop
further products of $\CQ$-closed 2-form operators multiplied by a
$\CQ$-exact operator, which will require contact
terms \cite{Moore:1997pc, LoNeSha}. Such cases are included implicitly
in the discussion in on the ``General $\CQ$-exact operator''.

We can give a closed expression for an arbitrary product of $\CQ$-exact two-form operators,
\be
\left\langle \prod_{j=1}^\ell  G_2^{(j)}(\bfx_j) \right\rangle,
\ee
with
\be
\label{G2j}
G_2^{(j)}=\sqrt{2} i\, \partial_{\bar a} V^{(j)}\, \eta\, \chi+ i
V^{(j)}\, (F_+-D),\qquad j=1,\dots,\ell.
\ee
We will prove below that the sum over fluxes can written as the
following total derivative
\be
\label{ellQexact}
\partial_{\bar \tau}\left( \left(\prod_{j=1}^\ell B(\bfx_j,
  J)\,V^{(j)}\right)\, \Psi_\bfmu^J[\CK_2^{(\ell)}]\right),
\ee
where the kernel $\CK^{(\ell)}_2$ is given by Equation (\ref{Kell}) in
terms of the Hermite polynomial $H_{\ell-1}$. The results of
Vign\'eras \cite{Vigneras:1977} imply that the integrand single valued
if the $V^{(j)}$ transform as $d\bar u/d\bar a$.

We have thus demonstrated that any product of $\CQ$-exact two-form observables can be
expressed as a total $\bar \tau$-derivative. The form of the kernel
ensures moreover that the integrand is well-defined, as long as the
functions $V^{(\ell)}$ transform with the same weight and multiplier system as $\frac{d\bar u}{d\bar a}$ under
$\Gamma^0(4)$ transformations.

\subsubsection*{4-form operators}
We can similarly treat the most general $\CQ$-exact 4-form
operator. If we leave aside terms which have an odd number of
fermionic fields and terms involving derivatives, the most general
operator has the form
\be \label{G4}
\begin{split}
G_4 & = \left\{ \CQ, V_{\chi,F_+}  \chi \wedge F_+ + V_{\chi,D}  \chi \wedge D  \right\}\\
&=\sqrt{2}i\partial_{\bar a}V_{\chi,F_+}\,\eta\chi\wedge F_++i\,V_{\chi,F_+}\,(F_+-D)\wedge F_+\\
&\quad + i V_{\chi,D} (F_+-D)\wedge D.
\end{split}
\ee
We aim to evaluate $\langle \int_M G_4 \rangle$. Using
(\ref{GaussDint}), (\ref{Dint2}) and
\be
\label{intsD}
\begin{split}
&\int dD\,\left[\int_M D\wedge D\right]\,e^{- \int_M \mathcal{L}_{0,D}}=-\frac{8\pi^2i}{y}\sqrt{\frac{2}{y}},
\end{split}
\ee
we find
\be
\begin{split}
&\int dD\, \int_M G_4 \,e^{- \int_M \mathcal{L}_{0,D}}= 2\pi i \sqrt{\frac{2}{y}}\left[ \sqrt{2}i\partial_{\bar a}V_{\chi,F_+}\, \int_M \eta\chi\wedge F_+\right.\\
& \left. \qquad +i\,V_{\chi,F_+}\,\int_M
  (F_+-i\frac{\sqrt{2}}{4y}\frac{d\bar \tau}{d\bar a}\,  \eta
  \chi)\wedge F_+ +i V_{\chi,D} \int_M (F_+\wedge i\frac{\sqrt{2}}{4y}\frac{d\bar \tau}{d\bar a}\eta \chi)+\frac{4\pi i}{y}V_{\chi,D}\right]
\end{split}
\ee
Integrating over the fermionic zero-modes gives the kernel $\CK_4$ for $\Psi_\bfmu^J$,
\be \label{K40}
\CK_{4}(\bfk) = \frac{8\sqrt{2}\pi^2 B(\bfk,J)}{\sqrt{y}} \left( \partial_{\bar \tau} V_{\chi,F_+} - \frac{i}{4y}V_{\chi,F_+} -\pi i V_{\chi,F_+} \bfk_+^2  \right).
\ee
Also $\Psi_\bfmu^J[\CK_{4}]$ can be expressed as an anti-holomorphic derivative in accordance to the previous subsection
\be
\Psi_\bfmu^J[\CK_{4}]=4
    \sqrt{2} i\partial_{\bar \tau} \left( V_{\chi,F_+}
  \Psi_{\bfmu}^J[\CK_{\rm{p}}]  \right).
\ee
with $\CK_{\rm p}$ as in (\ref{CK0}).

\subsubsection*{General $\CQ$-exact operator}
\label{GeneralCQ}
We have seen now a number of classes of $\CQ$-exact operators, whose
$u$-plane integrand can be expressed as a total derivative with
respect to $\bar \tau$. We will
demonstrate that this is not a coincidence but a generic
phenomenon. To this end, we reduce to the zero-mode sector
from the beginning and include the $\CQ$-exact part of the Lagrangian
in the observable. Recall that the zero-mode Lagrangian can be expressed as
\be
\CL_0=\frac{i}{8\pi} \tau F\wedge F+\left\{ \CQ,W \right\},
\ee
with $W=-\frac{i}{8\pi}\,y\,\chi\,(F_++D)$. We can rewrite
\be
\left\{\CQ, \CO\right\}\,e^{-\mathcal{L}_0}=\left\{\CQ, \widetilde
  \CO\right\}\,q^{-\bfk^2/2},
\ee
with $\widetilde \CO=\CO\,e^{-\int_M \left\{\CQ,W\right\}}$. This will
simplify the integrations over the fermion and auxiliary zero
modes.

To this end, let us expand $\widetilde \CO$ in terms of $\eta$ and
$\chi$, and integrate $\chi$ over a cycle $\bfx\in
H_2(M,\mathbb{Q})$, such that the operator $\widetilde \CO\in
H_0(M)$. The expansion then reads
\be
\widetilde \CO(\bfx)=\sum_{m=0,1}\widetilde \CO_{m,0}\,\eta^m+\sum_{m=0,1}\widetilde \CO_{m,1}\,\eta^m\,\int_\bfx \chi,
\ee
where $\widetilde \CO_{m,n}$ are functions of $a$, $\bar a$, $\int F$
and $\int D$.
With the $\CQ$-commutaters (\ref{barQcomm}) restricted to zero modes, we have
\be
\left\{\CQ, \widetilde \CO(\bfx)\right\}=\sqrt{2} i \partial_{\bar a}
\widetilde \CO_{0,0}\,\eta+\sqrt{2} i \partial_{\bar a}
\widetilde \CO_{0,1}\,\eta\,\int_\bfx \chi-i \int_\bfx (F_+-D)\sum_{m=0,1} \widetilde \CO_{m,1}\,\eta^m.
\ee
Only the term with $\widetilde \CO_{0,1}$ survives the integration over fermion zero modes,
\be
\int d\eta\, d\chi\,\left\{\CQ, \widetilde
  \CO(\bfx)\right\}=-\sqrt{2}i\, B(\bfx,J)\,\partial_{\bar a} \widetilde
\CO_{0,1},
\ee
where
\be
\widetilde \CO_{0,1}=\CO_{0,1}\,q^{-\bfk_-^2/2}\bar q^{\bfk_+^2/2}\, \exp\!\left(\frac{y}{8\pi} \int_M D^2\right).
\ee
We thus find that the $u$-plane integrand can be expressed as a
total $\bar \tau$-derivative for any $\CQ$-exact observable. Moreover,
the only term of $\widetilde
\CO$ which contributes to the integrand is linear in $\chi$ and independent of
$\eta$.

Let us consider a product $\prod_{j=1}^\ell G_2(\bfx_j)$ of
$\CQ$-exact operators. We can express this as $\{\CQ,\CO^{(\ell)}\}$, with
\be
\CO^{(\ell)}= V^{(1)}\, \int_{\bfx_1} \chi\, \times \prod_{j=2}^\ell G_2^{(j)}(\bfx_j),
\ee
with $G_2^{(j)}$ as in (\ref{G2j}). The coefficient of $\eta^0\,\int_{\bfx_1}\chi$, $\CO^{(\ell)}_{0,1}$, of this operator is
given by
\be
\CO_{0,1}^{(\ell)}=i^{\ell-1}B(F_+-D,J)^{\ell-1}\,V^{(1)}\,\prod_{j=2}^\ell
V^{(j)}\,B(\bfx_j,J).
\ee
Integrating out $D$ leads to expressions in terms of
the Hermite polynomials as suggested below Equation
(\ref{ellQexact}). To this end, recall the following integral
formula for the Hermite polynomials,
\be
\label{Hermite}
H_\ell(s)=\frac{2^\ell}{\sqrt{\pi}} \int^\infty_{-\infty} dt\,(s-it)^\ell\, e^{-t^2}.
\ee
The first few $H_\ell$ read
\be
\begin{split}
&H_0(s)=1,\\
&H_1(s)=2s,\\
&H_2(s)=4s^2-2.
\end{split}
\ee
Using the identity (\ref{Hermite}), we find
\be
\begin{split}
&\int dD\, \CO^{(\ell)}_{0,1}\,\exp\!\left(\frac{y}{8\pi} \int_M D^2\right)  = \\
&\qquad 2\sqrt{\pi}\left(i\sqrt{\frac{2\pi}{y}}
\right)^\ell\,H_{\ell-1}(\sqrt{2\pi y}\,B(\bfk,J))\,\prod_{j=1}^\ell
V^{(j)}\,B(\bfx_j,J).
\end{split}
\ee
We now arrive at Equation (\ref{ellQexact}) for the sum over fluxes,
with kernel $\CK^{(\ell)}_2$
\be
\label{Kell}
\CK^{(\ell)}_2=-2i\sqrt{2\pi}\left(i\sqrt{\frac{2\pi}{y}}
\right)^\ell\,H_{\ell-1}(\sqrt{2\pi y}\,B(\bfk,J)).
\ee

\subsection{A holomorphic self-dual operator}
\label{sectionIpm}
We have seen in the previous subsections examples of a holomorphic
operator combined with anti-selfdual field strength, and an
anti-holomorphic operator combined with a self-dual field
strength. The low energy expressions illustrate that they neatly
satisfy the constraints of the duality group. In this section, we
consider a $\CQ$-exact operator which is holomorphic in $a$ and
involves a self-dual field strength. We denote the UV operator by
$I_\diamond$, which reads explicitly
 \be \label{OP1}
I_\diamond(\bfx)=\int_{\bfx} \{{\CQ}, \text{Tr} [\phi\, \chi] \}.
 \ee
Since the integrand is not a descendant of $K$ or $L$, the IR operator
$\widetilde I_\diamond(\bfx)$ does not follow straightforwardly from the
UV expression. The discussion in this subsection is therefore of a more
speculative nature.

We take the following Ansatz for the IR observable
\be
I_\diamond(\bfx) \xrightarrow[\text{}]{\text{RG
    flow}} \widetilde I_\diamond(\bfx)=\int_\bfx \left( \frac{du}{da} \left( \, F_{+}-D\right)+g(\tau)\, \frac{du}{da}\, \left( \, F_{+}+D\right) \right),
\ee
where $g(\tau)$ is an unknown function, which we aim to fix below. From the UV definition, one would expect that $g$
vanishes, but we will see below that the integrand is not modular
invariant in that case. We will require that $g$ is a
non-perturbative correction, and vanishes exponentially fast in the
weak-coupling limit $\tau\to i\infty$. We will then demonstrate that
$g$ is uniquely determined by modularity.

Integration over the fermion zero modes after insertion of $\widetilde I_\diamond(\bfx)$, leads to the kernel
\be
\begin{split}
\CK_{\diamond}(\bfk)=&  \frac{i B(\bfx, J)}{2\sqrt{2y}} \,
\frac{du}{da} \, \left\{ \left( 4\pi \, \bfk_+^2 + \frac{1}{y} \right)-g(\tau) \left( 4\pi \, \bfk_+^2 - \frac{1}{y} \right)\right\}.
\end{split}
\ee

To satisfy the requirements that the integrand is $\Gamma^0(4)$ invariant and $g$
is non-perturba\-tive, we set
\be
g(\tau)=\frac{\pi}{6}\left(2E_2(\tau)-\vartheta_3(\tau)^4-\vartheta_4(\tau)^4 \right) =-16\pi\,q+O(q^2),
\ee
where $E_2$ is the Eisenstein series, and the $\vartheta_j$
are Jacobi theta series (\ref{Jacobitheta}). We set furthermore
\be
\widehat g(\tau)=\frac{\pi}{6}\left(2\widehat E_2(\tau)-\vartheta_3(\tau)^4-\vartheta_4(\tau)^4 \right),
\ee
with $\widehat{E}_2(\tau)$ the non-holomorphic Eisenstein series
$\widehat{E}_2(\tau) = E_2(\tau) - \frac{3}{\pi y }$. We see that
$\widehat g$ transforms as a weight two modular form of $\Gamma^0(4)$,
and that for $\tau\to i\infty$, the function $\widehat{g}(\tau)$
behaves as $-\frac{1}{y}+ O(q)$. We can now express $\Psi_\bfmu^J$
as a total derivative to $\bar \tau$
\be \label{Ker3}
\Psi_\bfmu^J[\CK_\diamond](\tau, \bar{\tau})=\frac{i}{2}
B(\bfx,J)\frac{du}{da}\, \frac{d}{d \bar{\tau}}\left(\sqrt{y}\, \widehat{g}(\tau)\, \Psi_\bfmu^J[1](\tau,\bar \tau)  \right),
\ee
where as before we have not included the term $\frac{d
  \bar{\tau}}{d\bar{a}}$, which is the Jacobian for the change of
variable to $\bar \tau$. One may verify that $\Psi_\bfmu^J[\CK_\diamond]$ has the same
transformation properties as $\Psi_\bfmu^J[\CK_+]$.

\subsection{Summary}
\label{CorrSum}
Let us give a summary of the results of this section. We have found
that vacuum expectation values of $\CQ$-exact operators can be
expressed as integrals whose integrands can be written as a total $\bar \tau$-derivative
after integration over the auxiliary field $D$ and the fermionic zero modes. The vev of a $\CQ$-exact operator takes therefore the form
\be
\label{SummCor}
\langle \left\{ \CQ,\CO \right\}\rangle=\int_{\BH/\Gamma^{0}(4)} d\tau\wedge d\bar
\tau \,\partial_{\bar \tau}\left( \widetilde \nu\,\, W_\CO\,\Psi_\bfmu^J[\CK_\CO] \right),
\ee
for some non-holomorphic function $W_\CO$ and kernel $\CK_\CO$, which
 both depend on $\CO$. Given the total derivative, we can easily
evaluate the integral using Stokes' theorem, which reduces the
integral to three arcs around the cusps of
$\mathbb{H}/\Gamma^0(4)$.

For a more standard treatment, we map the integral over
$\mathbb{H}/\Gamma^0(4)$ to an integral over
$\CF_\infty$, by mapping the six $\mathrm{SL}(2,\mathbb{Z})/\Gamma^0(4)$
images of $\CF_\infty$ in $\mathbb{H}/\Gamma^0(4)$ back to
$\CF_\infty$. Equation (\ref{SummCor}) can then be expressed as
\be
\langle \left\{ \CQ,\CO \right\}\rangle=\int_{\CF_\infty} d\tau\wedge d\bar
\tau \,\partial_{\bar \tau}F_\CO,
\ee
where $F_\CO$ is the sum of the six transformations of $\widetilde \nu\,\,
W_\CO\,\Psi_\bfmu^J[\CK_\CO]$ by the elements of $\mathrm{SL}(2,\mathbb{Z})/\Gamma^0(4)$. It has a $q$-expansion of the form
\be
F_\CO= y^{-s}\sum_{m,n} c(m,n)\,q^m\bar q^n,
\ee
or a finite sum of such terms with different $s$. For the $\CQ$-exact
operator $\widetilde I_+(\bfx)$, we have seen that
$m$ and $n$ can be both negative leading to a divergence for $\tau\to
i\infty$. Moreover, it is possible that $m=n<0$, for which the
standard renormalization does not apply. We introduce a new renormalization
prescription in the next section which also allows us to work with
operators leading to terms with $m=n<0$.

\section{Renormalization of modular integrals}
\label{regularization}

The previous section discussed the importance of integrals of the form
\be
\label{If}
\CI_f=\int_{\mathcal{F}} d\tau\wedge d\bar \tau \,y^{-s}\,f(\tau,\bar
\tau),
\ee
for supersymmetric field theories, where $f$ is a non-holomorphic modular form of weight $(2-s,2-s)$, and
$\CF$ a fundamental domain for the modular group, $\CF =
\BH/{\rm SL}(2,\BZ)$. We will discuss in this section the
evaluation, regularization and renormalization of integrals of this form, which has
been developed in the mathematical literature in the context of inner products for weakly holomorphic
modular forms \cite{1603.03056}.\footnote{A  weakly holomorphic modular form $f(\tau)$ is a modular form which is holomorphic on the interior of $\mathbb{H}$ but may diverge for $\tau\to i\infty \cup \mathbb{Q}$.}

\subsection{Renormalization of integrals over $\CF_\infty$}
\label{regintFinf}
We start by considering the integral over a single term $q^m\,\bar
q^n$ in the Fourier expansion of $f$.\footnote{We will justify in Section \ref{regmodint} that the Fourier series and the integral can
be exchanged.} To this end, consider the set $\CT$ of triples $(m,n,s)$, defined by
\be
\CT=\left\{m,n\in\mathbb{R}, s\in \mathbb{Z}/2\, \vert\,  m-n\in\mathbb{Z}\right\}.
\ee
For $(m,n,s)\in \CT$, we consider the integral
\be \label{Lmnr0}
L_{m,n,s}=\int_{\mathcal{F}_\infty} d\tau\wedge d\bar \tau \,y^{-s}\,q^m \bar
q^n,
\ee
where $\CF_\infty$ is the common keyhole fundamental domain
$\CF=\BH/\mathrm{SL}(2,\BZ)$ pictured in Figure
\ref{fund_domain}. Since $\CF_\infty$ is non-compact and the
integrand may diverge for $y\to \infty$, this is an improper integral.
It should be understood as the limiting value of integrals over compact
domains, which approach $\CF_\infty$.
To this end, we introduce the compact domain $\CF_Y$ by restricting $\mathrm{Im}(\tau)\leq Y$
for some $Y\geq 1$.\footnote{One
may consider a more general upperbound with $Y$ being a function of
$\mathrm{Re}(\tau)=x$. This will not affect the final result.} The
boundaries of $\CF_Y$ are given by the following arcs
\be
\label{arcs}
\begin{split}
&1:\quad \tau=\tfrac{1}{2}+iy,\qquad \,\,\,\, y\in
[\tfrac{1}{2}\sqrt{3},Y],\\
& 2:\quad \tau=x+iY,\qquad  \,\,\,\, x\in [-\tfrac{1}{2},\tfrac{1}{2}],\\
&3:\quad  \tau=-\tfrac{1}{2}+iy,\qquad y\in
[\tfrac{1}{2}\sqrt{3},Y],\\
&4: \quad  \tau=i\,e^{i\varphi},\qquad \,\,\,\,\,\,\,\,\,\,\,\, \varphi\in [-\tfrac{\pi}{6},\tfrac{\pi}{6}].\\
\end{split}
\ee
In the limit, $\lim_{Y\to
  \infty} \CF_Y$ we recover $\CF_{\infty}$. We then regularize $L_{m,n,s}$ as
\be
L_{m,n,s}(Y)=\int_{\mathcal{F}_Y} d\tau\wedge d\bar \tau \,y^{-s}\,q^m \bar
q^n,
\ee
for  $(m,n,s)\in\CT$, and define
\be
\label{LmnsLimit}
L_{m,n,s}=\lim_{Y\to \infty} L_{m,n,s}(Y),
\ee
provided the limit exists. To study the dependence on $Y$, we
split the compact domain $\CF_Y$ into $\CF_1$ plus a rectangle
$[-\frac{1}{2},\frac{1}{2}]\times [1,Y]$ as shown below in Figure
\ref{FD4}.
\begin{figure}[h]
\begin{center}
\includegraphics[scale=1.8]{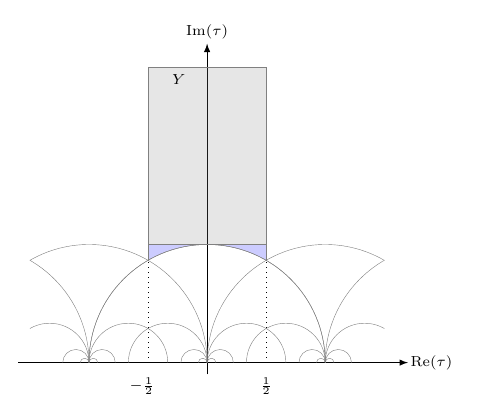}
\caption{Splitting of $\CF_Y$ into $\CF_1$ (the blue region) and the rectangle $\CR_Y$ (gray region). \label{FD4}}
\end{center}
\end{figure}
The split of $\CF_Y$, gives for $L_{m,n,s}(Y)$
\be \label{Lmnr1}
\begin{split}
L_{m,n,s}(Y)=&\int_{\mathcal{F}_1} d\tau\wedge d\bar \tau \,y^{-s}\,q^m \bar
q^n -2i\int_{-\frac{1}{2}}^{\frac{1}{2}} \int_{1}^Y
dx\wedge dy\, \,y^{-s}\,q^m \bar
q^n.
\end{split}
\ee
The first term on the right hand side is finite and independent of
$Y$. In the second term, we integrate
over $x$, which gives zero unless
$m=n$,
\be
\label{RY}
-2i\,\delta_{m,n} \int_{1}^Y dy\, \,y^{-s}\, e^{-4\pi m y}.
\ee
We thus find that $\lim_{Y\to\infty} L_{m,n,s}(Y)$ converges, except if
$m=n<0$, or if $m=n=0$ with $s\leq 1$. Let us denote this set by $\cD$,
\be
\label{DefcD}
\cD=\left\{ (m,n,s)\in\CT\,|\, m=n<0 \right\} \cup \left\{
  (0,0,s)\in\CT\,|\,  s\leq 1 \right\}.
\ee

The correlation functions discussed in Section \ref{PathCor} give rise to $(m,n,s)\in \cD$, suggesting that $\CQ$-exact observables appear to diverge rather than vanish. To resolve the tension of this divergence with the structure of topologically twisted theories, we aim to regularize and
renormalize such integrals. The cases with $m=n=0$ are renormalized in the standard way \cite{Dixon:1990pc, Harvey:1995fq, Borcherds:1996uda}: as the constant term of the integral for
sufficiently large $s$, which gives 0 for (\ref{RY}) if $s=1$ and otherwise $2i/(s-1)$. To treat the
cases with $m=n<0$, we put forward in this section a regularized and renormalized version $L^{\rm r}_{m,n,s}$, of $L_{m,n,s}$ for all $(m,n,s)\in \CT$.

Before introducing $L^{\rm r}_{m,n,s}$, let us note that the limit of the sum
\be
\lim_{Y\to\infty}\left[ L_{m,n,s}(Y)+2i\, \delta_{m,n} \int_{1}^Y dy\, \,y^{-s}\, e^{-4\pi my}\right]=L_{m,n,s}(1),
\ee
is finite. In the definition for $L^{\rm r}_{m,n,s}$, we will subtract from the two
terms in the brackets, an appropriately regularized counter part of the
second term. To this end, let us introduce the generalized exponential integral $E_\ell(z)$.
For $\mathrm{Re}(z)>0$, $E_\ell(z)$ is defined by
\be
\label{gen_exp}
E_\ell(z)=\int_{1}^\infty e^{-z\,t}t^{-\ell}dt.
\ee
Integral shifts of the parameter $\ell$ are related by partial integration
\be
\label{Ezpartial}
e^{-z}=z\,E_\ell(z)+\ell\,E_{\ell+1}(z).
\ee
We can also express $E_\ell(z)$ in terms of the incomplete Gamma function $\Gamma(k,z)$,
\be
\label{GammaE}
\Gamma(k,z)=\int_z^\infty e^{-t}\, t^{k-1} dt=z^{k} E_{1-k}(z).
\ee
With the analytic continuation of $\Gamma(k,z)$, we can extend the
domain of $E_\ell(z)$ to the full complex plane. We define
\be
\label{gen_exp2}
E_\ell(z) = \begin{cases}
\quad z^{\ell-1}{\displaystyle \int_z}^{\,\infty}
e^{-t}\,t^{-\ell}~dt,  &  \text{for  }  z\in \mathbb{C}^*,  \\
& \\
\quad \quad \quad \quad\displaystyle{\frac{1}{\ell-1}}, & \text{for  }   z=0,\,\, \ell\neq
1,\\
&\\
\quad \quad \quad \quad \quad 0, & \text{for  } z=0, \,\, \ell=1,
\end{cases}
\ee
where for non-integral $\ell$, we fix the branch of $t^{-\ell}$ by specifying
  that the argument of any complex number $\rho\in
\mathbb{C}^*$ is in the domain $(-\pi,\pi]$. For $s\in \mathbb{R}^+$,
we have $\mathrm{Im}(E_\ell(-s))=\pm
\frac{\pi\,s^{\ell-1}}{\Gamma(\ell)}$, where the sign is $+$ if the
contour $[-s,\infty)$ is deformed to the lower half plane of the complex half-plane to
avoid the singularity at $t=0$, and the sign is $-$ if the contour is
deformed to the upper half-plane. 

 In terms of this function
$E_\ell(z)$, we finally define $L^{\rm r}_{m,n,s}$ for all $(m,n,s)\in\CT$:
\be
\begin{split}
\label{Lmnreg}
L^{\rm r}_{m,n,s}&= \int_{\mathcal{F}_1} d\tau\wedge d\bar \tau \,y^{-s}\,q^m\bar q^n-2i\,\delta_{m,n}\,E_{s}(4\pi m),
\end{split}
\ee
which regularizes and renormalizes the ill-defined $L_{m,n,s}$.

\subsection{Modular invariant integrands}
\label{regmodint}
We provide in this subsection the  prescription to renormalize $\CI_f$.
Let us start with the integral of a modular form over the fundamental domain,
\be
\label{If}
\CI_f=\int_{\mathcal{F}_\infty} d\tau\wedge d\bar \tau \,y^{-s}\,f(\tau,\bar
\tau),
\ee
where $f(\tau,\bar \tau)$ is a non-holomorphic modular form  for
${\rm SL}(2,\mathbb{Z})$ of weight
$(2-s,2-s)$, with Fourier expansion
\be
\label{expansion_f}
f(\tau,\bar\tau)=\sum_{m,n\gg -\infty}\, c(m,n)\, q^m \bar
q^n,
\ee
where the $c(m,n)$ are only non-zero if $m-n \in \mathbb{Z}$ by the requirement
that $f$ is a modular form. We assume that $f$ is in fact a function on
$\mathbb{H}\times \mathbb{\bar H}$, which satisfies
\be
\label{ftrafo}
f\!\left(
  \frac{a\tau+b}{c\tau+d},\frac{a\sigma+b}{c\sigma+d}\right)=(c\tau+d)^{2-s}
(c\sigma+d)^{2-s} f(\tau,\sigma),
\ee
where for $s\in
  \mathbb{Z}+\tfrac{1}{2}$, we specify the branch of the square root
  by requiring that the argument of $c\tau+d$ is in $(-\pi,\pi]$. For
  a single factor $(c\tau+d)^{2-s}$, consistency of the square root
  and $\operatorname{SL}(2,\mathbb{Z})$ requires a non-trivial multiplier system. For
  $f(\tau,\sigma)$, the multiplier systems for $\tau$ and $\sigma$ are complex conjugate and
  multiply to 1 on the rhs of (\ref{ftrafo}).

For the physical correlation functions of Section \ref{PathCor}, we
have to allow $f$ with a finite number of polar terms, i.e. there is
an $M\in \mathbb{Z}$ such that $c(m,n)=0$ if $m<M$ or $n<M$, such that
the number of terms with $m+n<0$ is finite. For sufficiently large $m$
and $n$, double application of the well-known saddle point argument shows that the coefficients $c(m,n)$ are bounded by
\be
\label{boundcmn}
c(m,n) < e^{\sqrt{Km}\,+\,\sqrt{K n}},
\ee
for some constant $K>0$. The sum over $m$ and $n$ is therefore
absolutely convergent for $\mathrm{Im}(\tau)<\infty$.

Due to the terms with $m+n\leq 0$, the integrand in (\ref{If}) diverges for
$y\to \infty$, such that the integral is ill-defined. If there are no terms with $m=n<0$, the integral is defined
using a well-known regularization \cite{Dixon:1990pc, Harvey:1995fq,
  Borcherds:1996uda}, but we have seen in Section \ref{PathCor} also
terms with $m=n<0$ may appear in correlation functions on the Coulomb
branch. To regularize these integrals, we introduce a cut-off $Y$ for
$\mathrm{Im}(\tau)$ as in Subsection \ref{regintFinf}, and define the
integral $\CI_f(Y)$ of $f$ over this
domain $\CF_Y$ (\ref{arcs}),
\be
\CI_f(Y)=\int_{\mathcal{F}_Y} d\tau\wedge d\bar \tau
\,y^{-s}\,f(\tau,\bar \tau).
\ee

We regularize the divergence of $\CI_f(Y)$ by subtracting terms
involving the generalized exponential function $E_s(z)$ defined in
(\ref{gen_exp2}). More precisely, we replace $\CI_f$ by its
regularized and renormalized version $\CI^{\mathrm{r}}_f$, defined as
\be
\begin{split} \label{Irf}
\CI^{\rm r}_f= &\lim_{Y\to \infty} \left[ \CI_f(Y)-2i \sum_{m \gg -\infty} c(m,m)\, Y^{1-s} E_s(4\pi m Y) \right].
\end{split}
\ee

Let us verify that the limit is well-defined. Since the
domain $\CF_Y$ is compact and the sum over $m$ and $n$ is absolutely
convergent on $\CF_Y$, we can exchange the double integral and the sum. Thus,
\be
\CI_f(Y)=\sum_{m,n\gg -\infty} c(m,n)\, L_{m,n,s}(Y),
\ee
with $L_{m,n,s}(Y)$ as in (\ref{Lmnr1}). We substitute this
expression in (\ref{Irf}). Using
$$\int_1^Y dy\, y^{-s}\, e^{-4\pi m y} = E_s(4\pi m) - Y^{1-s} E_s(4\pi m Y),$$
we arrive at
\be
\CI_f^{\rm r}=\sum_{m,n\gg -\infty} c(m,n)\, L^{\rm r}_{m,n,s}\,,
\ee
with $L^{\rm r}_{m,n,s}$ as in (\ref{Lmnreg}).
This is finite since there are at most a finite number of terms with
$m=n<0$, and the sum over the other $m$ and $n$ is absolutely
convergent.

\subsection{Evaluation using Stokes' theorem}
\label{StokesThm}
If we assume that the integrand can be expressed as a total
derivative with respect to $\bar \tau$, we can evaluate the integral using Stokes' theorem, and
we will find that $\CI_f^{\rm r}$ takes an elegant form in this case. To this end, let us
write $y^{-s}f(\tau,\bar \tau)$ as
\be
\label{dh=f}
\partial_{\bar \tau} \widehat h(\tau,\bar \tau)=y^{-s} f(\tau,\bar \tau),
\ee
such that the integrand of (\ref{If}) is in fact exact and equal to $-d(d\tau\,\widehat h)$.
Note that this does not imply that $d\bar \tau\,\partial_{\bar
  \tau}\widehat h$ is exact, since $d \widehat h=d\tau\,\partial_{\tau}\widehat h+d\bar \tau\,\partial_{\bar
  \tau}\widehat h$. For our application to modular
integrals, $\widehat h(\tau,\bar \tau)$ transforms as a modular form
of weight two. Eq. (\ref{dh=f}) can be integrated using
$E_\ell(z)$. For $s\neq 1$,\footnote{We follow here the
  convention for Maass forms as in \cite{bruinier2004}. In other
  literature on Maass forms such as \cite{1603.03056}, $E_\ell(s)$
  is sometimes replaced by the function $s^{\ell-1} W_\ell(-s/2)=\mathrm{Re}(E_\ell(s))$, $(s\neq 0)$. This has no effect for $s>0$, but terms with $s<0$ lead to
  additional contributions involving ${\rm Im}(E_\ell(s))$ in the final result
  for $\CI^{\rm r}_f$ (\ref{Constant}). Ref. \cite[Definition 3.1]{1603.03056}
  corrected for this in the definition of their inner-product.}
\be
\label{widehathconv}
\widehat h(\tau,\bar\tau)=h(\tau) + 2i\,y^{1-s} \sum_{m,n \gg -\infty} c(m,n)\,q^{m-n}E_s(4\pi n y),
\ee
while for $s=1$, the terms with $n=0$ in the sum should be replaced by
$$-2i\,\log(y)\sum_{m\gg -\infty} c(m,0)\,q^m.$$
The $c(m,n)$ in (\ref{widehathconv}) are the Fourier coefficients of $f$ (\ref{expansion_f}), and $h$ is a (weakly) holomorphic function with Fourier expansion
\be
\label{holoh}
h(\tau)=\sum_{m\gg -\infty \atop m\in \mathbb{Z}} d(m)\,q^m.
\ee
Since there are no holomorphic modular forms of weight two for
$\mathrm{SL}(2,\BZ)$, $h(\tau)$ is uniquely determined by the
coefficients $d(m)$ with $m<0$. However, since the $d(m)$, $m<0$, are
not determined by the  $c(m,n)$,  the space of weakly holomorphic modular forms of
weight 2 gives an ambiguity in $h(\tau)$. We will discuss below (\ref{Constant}), that the integral
$\CI^{\mathrm r}_f$ is independent of this ambiguity.

Note that if $f=f(\bar \tau)$ is a (weakly) anti-holomorphic,
$\widehat h(\tau,\bar \tau)$ is annihilated by the weight $s$
hyperbolic Laplacian, and in this case almost satisfies the
requirements for a harmonic Maass form
\cite{Larry}.\footnote{A harmonic Maass form of weight $k$ is
  annihilated by the weight $k$ hyperbolic Laplacian, whereas the
  weight of $\widehat h(\tau,\bar \tau)$ is 2 independently of $s$.} Moreover, if $f$ is
anti-holomorphic, $h(\tau)$ is a mock modular form with shadow
$\overline f$ \cite{ZwegersThesis, MR2605321}.

The modular properties of $\widehat h(\tau,\bar \tau)$ imply
interesting transformations for $h(\tau)$. Let us consider this for the
case that $f$ depends on both $\tau$ and $\bar \tau$, but is such
that the $c(m,n)$ in (\ref{widehathconv}) are only non-vanishing for
$n> 0$ (or $n\geq 0$ and $s>1$). We can then express $\widehat h$ as
\be
\label{widehath}
\widehat h(\tau,\bar\tau)=h(\tau) + 2^s  \int_{-\bar \tau}^{i\infty}
\frac{f(\tau,-v)}{(-i(v+\tau))^s}dv.
\ee
Note that the two terms on the right hand side are separately
invariant under $\tau\to \tau+1$, while the transformation of
the integral under $\tau\to -1/\tau$ implies for $h(\tau)$,
\be
h(-1/\tau)=\tau^2 \left( h(\tau)+2^s \int_{0}^{i\infty} \frac{f(\tau,-v)}{(-i(v+\tau))^s}\,dv\right).
\ee

Let us return now to the generic case with $f(\tau,\bar \tau)$ of the form
(\ref{expansion_f}) and evaluate $\CI^{\rm r}_f$. The integral over
$\CF_Y$ can then be carried out using Stokes' theorem, which reduces to a contribution from  the interval
$[-\frac{1}{2}+iY,\frac{1}{2}+iY]$. We thus find that the integral $\CI_f(Y)$ in (\ref{Irf}) equals for $s\neq 1$,
\be
\label{Stokesd0}
d(0) + 2i \sum_{m\gg -\infty} Y^{1-s}\,c(m,m)\,
E_s(4\pi m Y),
\ee
using expression (\ref{widehathconv}) for $\widehat h$. For $s=1$, we
apply the renormalization by analytic continuation in $s$ mentioned
below (\ref{DefcD}), which gives the same result.

The last step is to combine (\ref{Stokesd0}) with the other term in
Equation (\ref{Irf}), which gives
\be
\begin{split} \label{Constant}
\CI^{\rm r}_f&= d(0).
\end{split}
\ee
As a result the only contribution to the integral arises from the
constant term of $h(\tau)$. This obviously reduces to the standard renormalization for $\CI_f$ if
either $m$ or $n$ is non-negative \cite{Moore:1997pc,
  Harvey:1995fq}. We mentioned below Equation (\ref{holoh}), that
there is an ambiguity in $h$ due to the possibility to add a weakly
holomorphic modular form of weight two. Since the constant terms of
such modular forms vanishes, the result (\ref{Constant}) does not
depend on this ambiguity.

To see that the constant terms of such modular forms vanishes, let
$C(\tau)$ be a weakly holomorphic modular form of weight two. Since the first
cohomology of $\CF_\infty$ is trivial, the one-form
$C(\tau)\,d\tau$ is necessarily exact. The period
$\int^{Y+1}_{Y}C(\tau)\,d\tau$ therefore vanishes, which implies that its
constant term vanishes. Indeed, a basis of weakly
holomorphic modular forms of weight 2 is given by derivatives of
powers of the modular invariant $J$-function, $\partial_{\tau}\!
\left(J(\tau)^\ell\right)$, $\ell \in \mathbb{N}$, which have all
vanishing constant terms.

\section{Evaluation of correlation functions of $\CQ$-exact observables}
\label{evaluation}
We return to the $u$-plane integrals for correlation functions of
$\CQ$-exact observables $\bra \{\CQ,\CO\}  \ket$, where $\{\CQ,\CO\}$
may be a product of $\CQ$-exact and $\CQ$-closed operators as discussed
in Section \ref{PathCor}. As discussed in
Subsection \ref{GeneralCQ}, the corresponding $u$-plane integrals take
the form of a total $\bar \tau$-derivative for $\CQ$-exact
observables. This is the key property for their evaluation, and we can
therefore treat all such correlation function simultaneously as
indicated in Section \ref{CorrSum}.

Using the regularization of Section \ref{regularization}, we will show that the correlation functions of the form $\bra \{\CQ,\CO\}  \ket$ vanish, confirming the Ward-Takahashi
identities of the BRST symmetry. At this point recall from Subsection \ref{CorrSum}, that $\bra \{\CQ,\CO\}  \ket$
can be expressed as
\be \label{FourierCoeff}
\begin{split}
\langle \{\CQ,\CO\} \rangle=& \int_{\CF_{\infty}} d\tau \wedge
d\bar{\tau}~ \partial_{\bar \tau} F_\CO,
\end{split}
\ee
with
\be
F_{\CO}(\tau,\bar \tau)= y^{-s} \sum_{m,n} c(m,n)\,q^m\bar q^{n},
\ee
where only a finite number of $c(m,n)\neq 0$ for $m+n<0$. Let us first
evaluate (\ref{FourierCoeff}) using Section \ref{StokesThm}. Since
\be
\partial_{\bar \tau} F_\CO=-i\, y^{-s}\,\sum_{m,n}
c(m,n)\,(2\pi\,n+\tfrac{1}{2}\,s\,y^{-1})\, q^m\bar q^n,
\ee
we can identify $F_\CO$ with $\widehat h_1+\widehat h_2$ following (\ref{dh=f}).
Here $\widehat h_1$ is of the form (\ref{widehathconv}) and $\widehat
h_2$ as well, but with $s$ replaced by $s+1$. $F_\CO$ is a
(non-holomorphic) modular form of weight 2, and the discussion in
Section \ref{PathCor} did not include a holomorphic function
$h_1+h_2$. Indeed, since $F_\CO$ is a modular form of weight 2,
vanishing of $h_1+h_2$ is consistent with the modular properties. The
sum of constant terms $d_1(0)+d_2(0)$ thus vanishes, which demonstrates that $\langle \{\CQ,\CO\}
\rangle$ vanishes.

Alternatively, one may start from (\ref{Irf}) with $f=\partial_{\bar
  \tau} F_\CO$, such that $\langle \{\CQ,\CO\}
\rangle$ reads
\be
\label{QOreg}
\begin{split}
\langle \{\CQ,\CO\} \rangle=&\lim_{Y\to \infty}\left[\int_{\mathcal{F}_Y} d\tau\wedge d\bar \tau
\,\partial_{\bar \tau} F_\CO \right. \\
&\left.-2Y^{-s}\,\sum_{m\gg -\infty} c(m,m) (2\pi\,m\,
Y\,E_s(4\pi mY) +\frac{s}{2}\,E_{s+1}(4\pi mY)
  )\right].
\end{split}
\ee
To evaluate the integral over $\CF_Y$, we use Stokes' theorem. Modular
invariance of the integrand implies that only the arc at
$\mathrm{Im}(\tau)=Y$ contributes. Using (\ref{Ezpartial}) for the
second line, we arrive again at the desired result
\be
\begin{split}
\langle \{\CQ,\CO\}  \rangle&=
\sum_{m}c(m,m) \lim_{Y\to \infty} \left[ Y^{-s}\, e^{-4\pi
    Ym} - Y^{-s}\, e^{-4\pi
    Ym} \right]\\
&=0.
\end{split}
\ee
We have thus demonstrated that the correlation function of a generic
$\CQ$-exact observable vanishes with the current prescription.

Given that the vev of any $\CQ$-exact observable vanishes, power series of
$\CQ$-exact observables vanish as well. We have in particular
$$
\langle (1-e^{\alpha \{\CQ,\CO\}})\,\CO'\rangle =0,
$$
for arbitrary $\alpha \in \mathbb{C}$ and assuming that $\CO'$ is
$\CQ$-closed.  We can therefore safely add $\CQ$-exact terms to the
action. This justifies the inclusion of $e^{\widetilde I_+(\bfx)}$ in
the $u$-plane integrand as in \cite{Korpas:2017qdo}. It was, in fact,
precisely this question which motivated the present article.

\section{Discussion and conclusion}
We have revisited the evaluation of correlation
functions on the Coulomb branch of Donaldson-Witten theory. While vanishing of correlation
functions of $\CQ$-exact observables is important for the topological nature of the
theory, we have seen there are natural $\CQ$-exact observables whose
correlation functions appear to diverge due to contributions from the
boundary of field space. The divergences become most manifest after a change of
variables from $u$ to the complexified coupling constant $\tau\in
\BH/\Gamma^0(4)$. Depending on the observable, the integrand may
contain terms $q^m \bar{q}^n$ with $m,n$ both negative (where
$q=e^{2\pi i \tau}$), which diverge for $\tau\to \infty$.

We have demonstrated that such divergences can be cured using a new
prescription to regularize and renormalize the integrals over modular
fundamental domains. This prescription employs the analytic
continuation of the incomplete Gamma function, and was recently developed
for for the definition of regularized inner products of weakly holomorphic
modular forms \cite{1603.03056}. Strikingly, this results in a
vanishing expectation value for the correlation functions of
$\CQ$-exact observables in Donaldson-Witten
theory, confirming its BRST symmetry. With the new regularization we have demonstrated that all valid
$\CQ$-exact observables decouple from the $\CQ$-closed operators. A
central aspect of our analysis was that $\CQ$-exact observables lead
to a $u$-plane integrand which is a total derivative with respect to
$\bar \tau$. We will further elaborate on this aspect for $\CQ$-closed
observables in upcoming work \cite{to_appear}.

As we have restricted our analysis to Donaldson-Witten theory and four-manifolds
with $b_2^+=1$, there are immediate directions for future work. We
plan to analyze  in future work the BRST symmetry of other twisted theories
including those with matter and with superconformal symmetry. We would like to extend our discussion
also to four-manifolds with $b_2^+=0$, where one-loop determinants contribute
in addition to the zero modes.

Besides the $\CQ$-closed observables, the new prescription also
renormalizes correlation functions of observables outside the
$\CQ$-cohomology, which are ``unphysical'' from the point of view of the
topological theory. An example is $\langle \mathrm{Tr}[\bar \phi^2]\rangle$. We leave it for future work to see
whether such correlation functions may contain interesting information.

Another potential area of applications are string amplitudes; the
context in which previous regularizations were developed
\cite{Dixon:1990pc, Harvey:1995fq}. In particular, it is a standard result that 
the one-loop contribution $\CA_{1-\mathrm{loop}}$ to the vacuum energy in the bosonic string is divergent 
due to the presence of a tachyon \cite{Polchinski:1998rq}. Curiously, the new 
prescription gives a definite \underline{finite} value for this
amplitude! Recall that $\CA_{1-\mathrm{loop}}=i\,\CI_f$ with $f(\tau,\bar \tau)=|\eta(\tau)|^{-48}$. We
find for the value $\CA^\mathrm{r}_{1-\mathrm{loop}}$ after regularization
 \be
 \label{torus_amplitude}
\begin{split}
\CA^\mathrm{r}_{1-\mathrm{loop}}&=i\frac{2^{27}\,\pi^{14}}{\Gamma(14)}+
\sum_{m,n\geq -1} c(m,n)\,\mathrm{Re}(L^{\mathrm{r}}_{m,n,14})\\
&= i\,196\, 620.04\dots\,+\,58\,798.14\dots\,,
\end{split} 
 \ee
where we used
$\mathrm{Im}(E_\ell(-s))=+\frac{\pi\,s^{\ell-1}}{\Gamma(\ell)}$ for
$s>0$ since the imaginary part of the amplitude is naturally
positive.\footnote{Note added 20 March 2023: After \cite{Eberhardt:2023xck} appeared, a numerical error was found in
the Mathematica code to evaluate the formula. The corrected numerical
value (\ref{torus_amplitude}) agrees with the value in \cite[Section
8]{Eberhardt:2023xck} which employed the $i\varepsilon$ prescription put forward in
\cite{Witten:2013pra}. We thank Lorenz Eberhard for discussions.} Note
that the tachyon gives rise to the imaginary part of the amplitude. What, if 
any, are the physical consequences of this mathematical fact is an interesting 
open question.

\subsection*{Acknowledgments}
\vspace{-.2cm}
We thank Samson Shatashvili for discussions. GK would like to thank the Stanford Institute for Theoretical Physics and King's College London Department of Mathematics for hospitality. JM thanks the New High Energy Theory Center, Rutgers University for hospitality.
GM thanks the Stanford Institute for Theoretical Physics
for hospitality during the completion of this work. JM is supported by Laureate Award 15175 of the Irish Research Council. GM and IN are supported by the US Department of Energy under grant DE-SC0010008.

 %%%%%%%%%%%%%%%%%%%%%%%%%%%%%%%%%%%%%%%%%%%%%%
%%%%%%%%%%%%%%%%%%%%%%%%%%%%%%%%%%%%%%%%%%%%%%
%%%%%%%%%%%%%%%%%%%%%%%%%%%%%%%%%%%%%%%%%%%%%%
%%%%%%%%%%%%%%%%%%%%%%%%%%%%%%%%%%%%%%%%%%%%%%
%%%%%%%%%%%%%%%%%%%%%%%%%%%%%%%%%%%%%%%%%%%%%%

\appendix

%%%%%%%%%%%%%%%%%%

\section{Modular forms and theta functions}
\label{app_mod_forms}
We collect a few aspects of the theory of
modular forms and Siegel-Narain theta functions. See for more comprehensive treatments for example \cite{Serre,Zagier92,Bruinier08}.

\subsection*{Modular groups}
The modular group $\operatorname{SL}(2,\mathbb{Z})$ is the group of integer matrices
with unit determinant:
\be
\operatorname{SL}(2,\mathbb{Z})=\left\{ \left. \left( \begin{array}{cc}
a & b   \\
c & d  \end{array} \right) \right| a,b,c,d\in \BZ; \, ad-bc=1\right\}.
\ee
The congruence subgroup $\Gamma^0(n)$ is defined as:
\begin{equation}
\label{Gamma04}
\Gamma^0(n) = \left\{  \left.\left( \begin{array}{cc}
a & b   \\
c & d  \end{array} \right) \in \text{SL}(2,\BZ) \right| b = 0 \text{ mod } n \right\}.
\end{equation}

\subsection*{Jacobi theta functions}
The four Jacobi theta functions $\vartheta_j:\mathbb{H}\times
\mathbb{C}\to \mathbb{C}$, $j=1,\dots,4$, are defined as
\be
\label{Jacobitheta}
\begin{split}
&\vartheta_1(\tau,v)=i \sum_{r\in
  \mathbb{Z}+\frac12}(-1)^{r-\frac12}q^{r^2/2}e^{2\pi i
  rv}, \\
&\vartheta_2(\tau,v)= \sum_{r\in
  \mathbb{Z}+\frac12}q^{r^2/2}e^{2\pi i
  rv},\\
&\vartheta_3(\tau,v)= \sum_{n\in
  \mathbb{Z}}q^{n^2/2}e^{2\pi i
  n v},\\
&\vartheta_4(\tau,v)= \sum_{n\in
  \mathbb{Z}} (-1)^nq^{n^2/2}e^{2\pi i
  n v}.
\end{split}
\ee

We let $\vartheta_j(\tau,0)=\vartheta_j(\tau)$ for $j=2,3,4$.
Their transformations under the generators of $\Gamma^0(4)$ are
\be
\label{Jacobitheta_trafos}
\begin{split}
&\vartheta_2(\tau+4)=-\vartheta_2(\tau),\qquad
\vartheta_2\!\left(\frac{\tau}{\tau+1}\right)=\sqrt{\tau+1}\,\vartheta_3(\tau),  \\
&\vartheta_3(\tau+4)=\vartheta_3(\tau),\qquad
\vartheta_3\!\left(\frac{\tau}{\tau+1}\right)=\sqrt{\tau+1}\,\vartheta_2(\tau),  \\
&\vartheta_4(\tau+4)=\vartheta_4(\tau),\qquad
\vartheta_4\!\left(\frac{\tau}{\tau+1}\right)=e^{-\frac{\pi
    i}{4}}\sqrt{\tau+1}\,\vartheta_4(\tau).  \\
\end{split}
\ee
In particular, from the above we see that $\vartheta_{1}(\tau)$ gives a one dimensional representation of ${\rm SL}(2,\BZ)$ while $\vartheta_{i}(\tau)$ for $i=2,3,4$ give a three dimensional representation of ${\rm SL}(2,\BZ)$.

\section{Siegel-Narain theta function}
\label{SNtheta}
Siegel-Narain theta functions form a large class of theta functions of
which the Jacobi theta functions are a special case. For our
applications in the main text, it is sufficient to consider
Siegel-Narain theta functions for which the associated lattice
$\Lambda$ is a uni-modular lattice with signature
$(1,n-1)$ (or a Lorentzian lattice). We denote the bilinear form by
$B(\bfx,\bfy)$ and the quadratic form $B(\bfx,\bfx)\equiv Q(\bfx)\equiv \bfx^2 $.
Let $K$ be a characteristic vector of $\Lambda$, such that
$ Q(\bfk) + B(\bfk, K) \in 2\mathbb{Z}$ for each $\bfk\in \Lambda$.

Given an element $J\in \Lambda\otimes \mathbb{R}$ with $Q(J)=1$, we
may decompose the space $\Lambda\otimes \mathbb{R}$ in a positive
definite subspace $\Lambda_+$ spanned by $J$, and a negative definite
subspace $\Lambda_-$, orthogonal to $\Lambda_+$. The projections of a
vector $\bfk\in \Lambda$ to $\Lambda_+$ and $\Lambda_-$ are then given by
\be
\label{k+k-}
\bfk_+=B(\bfk,J)\, J, \qquad \qquad \bfk_{-} = \bfk-\bfk_+.
\ee

Given this notation, we can introduce the Siegel-Narain theta function of our
interest $\Psi^J_\bfmu[\CK]:\mathbb{H}\to \BC$, as
\be
\label{PsiJ}
\begin{split}
\Psi^J_\bfmu[\CK](\tau,\bar \tau)=&\sum_{\bfk\in
  \Lambda + \bfmu} \CK(\bfk)\,(-1)^{B(\bfk, K)} q^{-\bfk_-^2/2} \bar q^{\bfk_+^2/2}, \\
\end{split}
\ee
where $\bfmu\in \Lambda/2$ and $\CK: \Lambda \to \mathbb{C}$ is a
summation kernel. We also introduce the theta function including an
elliptic variable $\bfz$,
\be
\label{PsiJell}
\begin{split}
\Psi^J_\bfmu[\CK](\tau,\bar \tau,\bfz, \bar \bfz)=&\sum_{\bfk\in
  \Lambda + \bfmu} \CK(\bfk)\,(-1)^{B(\bfk, K)} q^{-\bfk_-^2/2} \bar q^{\bfk_+^2/2} \\
  & \times \exp\left( -2\pi i B(\bfz, \bfk_-) - 2\pi i B(\bar \bfz, \bfk_+) \right).
\end{split}
\ee

The modular properties of $\Psi^J_\bfmu[\CK]$ depend on $\CK$. For
$\CK=1$ and $\Psi^J_\bfmu[1]$, the modular transformations under the ${\rm SL}(2,\BZ)$ generators are
\be
\begin{split} \label{SL2Z}
\Psi^J_{\bfmu+K/2}[1](\tau+1,\bar{\tau}+1, \bfz, \bar{\bfz}) &= e^{\pi i (\bfmu^2-K^2/4)} \Psi_{\bfmu+K/2}[1](\tau,\bar \tau, \bfz + \bfmu, \bar{\bfz} + \bfmu), \\
\Psi^J_{\bfmu + K/2}[1]\left(-1/\tau, -1/\bar{\tau},
 \bfz/\tau, \bar{\bfz}/\bar \tau\right) &=
(-i\tau)^{\frac{n-1}{2}}(i\bar{\tau})^\frac{1}{2} \exp(-\pi i
\bfz^2/\tau+\pi i K^2/2)  \\
 & \qquad \times  (-1)^{B(\bfmu, K)}\,\Psi^J_{K/2}[1](\tau, \bar{\tau}, \bfz-\bfmu,\bar \bfz -\bfmu).
\end{split}
\ee
For the case of the partition function in Section \ref{path_int_Don}, we set the elliptic variables
$\bfz, \bar \bfz$ to zero. Using the above ${\rm SL}(2,\BZ)$
transformations and Poisson resummation one verify that
$\Psi^J_\bfmu[1]$
is a modular form for the congruence subgroup $\Gamma^0(4)$. The
transformations under the generators of this group read
\be
\label{Psitrafos}
\begin{split}
&\Psi^J_\bfmu[1]\!\left( \frac{\tau}{\tau+1},
  \frac{\bar \tau}{\bar \tau+1} \right)=(\tau+1)^{\frac{n-1}{2}}(\bar
\tau+1)^\frac{1}{2}\exp\!\left(\tfrac{\pi i}{4}K^2\right)
\Psi^J_{\bfmu}[1](\tau,\bar \tau),\\
&\Psi^J_\bfmu[1](\tau+4,\bar \tau +4)=e^{2\pi i
  B(\bfmu,K)}\,\Psi_\bfmu[1](\tau,\bar \tau),
\end{split}
\ee
where we have set $\bfz= \bar \bfz = 0$.
Transformations for other kernels appearing in the main text are
easily determined from these expressions.

\section{The self-dual twisted operator} \label{Twisting}
We discuss in this appendix the twisted supersymmetry generators
$\CQ$, $K$ and $L$, and we give a formula for $\{\CQ, L\}$ for an
arbitrary K\"ahler surface. Recall the global bosonic symmetry group of our theory $G=\mathrm{SU}(2)_-
\times \mathrm{SU}(2)_+ \times \mathrm{SU}(2)_R \times
\mathrm{U}(1)_R$. The first two factors correspond to the global
``Lorentz" rotations while the latter two factors correspond to the
$R$-symmetry.

The supersymmety generators $Q_{\alpha A}, \bar{Q}_{\dot{\alpha}}^{\,\, B}$,
written explicitly,  have the following non-zero anticommutator for a
local patch given by coordinates $x^m$ such that $m,n=0,\ldots,3$
\be
\label{SusyAlg}
\begin{split}
\{ Q_{\alpha A}, \bar{Q}_{\dot{\alpha}}^B \} &= 2\, \delta_{A}^{\,\, B} (\sigma^{m})_{\alpha\dot{\alpha}}\, P_{m}, \\
\{ Q_{\alpha A}, {Q}_{\beta B} \}  &= 2 \sqrt{2}\,\epsilon_{\alpha \beta}\,Z_{AB}, \
\end{split}
\ee
with $Z$ the central charge, $P_{m} \equiv \partial_{m}$ is the generator of translations, and $\sigma_m$ the Pauli matrices
\[
  \sigma_0 =  \left( {\begin{array}{cc}
   1 & 0 \\
   0 & 1 \\
  \end{array} } \right), \hspace{1em}
\sigma_1 =  \left( {\begin{array}{cc}
   0 & 1 \\
   1 & 0 \\
  \end{array} } \right), \hspace{1em}
 \sigma_2 =  \left( {\begin{array}{cc}
   0 & -i \\
   i & 0 \\
  \end{array} } \right), \hspace{1em}
  \sigma_3 =  \left( {\begin{array}{cc}
   1 & 0 \\
   0 & -1 \\
  \end{array} } \right).
\]
The $\alpha,\dot{\alpha}=1,2$ are indices of $\mathrm{SU}(2)_-$ and
$\mathrm{SU}(2)_+$ respectively. We define furthermore
\begin{eqnarray}
\sigma_{mn} &=& \frac{1}{4}(\sigma_{m} \bar{\sigma}_{n} - \sigma_{n} \bar{\sigma}_{m}), \
\end{eqnarray}
with $\bar{\sigma}_m$ the complex conjugate of $\sigma_m$.

Topological twisting amounts to redefining the spins of the fields of the vector multiplet and eventually allows to formulate a supersymmetric theory on a compact four-manifold. Our supercharges transform in the ${(\bf{1}, \bf{2}, \bf{2})}^{1} \oplus {(\bf{2},\bf{1},\bf{2})}^{-1}$ representation under the global group $G$. Originally, the rotation group is $K' = \mathrm{SU}(2)_- \times \mathrm{SU}(2)_+$ in the untwisted theory. The twist redefines the rotation group of the theory. There are two choices (related by conjugation)
\begin{enumerate}[(i)]
\item $K_1' =  \text{diag}(\mathrm{SU}(2)_- \times \mathrm{SU}(2)_R)\times \mathrm{SU} (2)_+ $,
\item $K_2' =  \text{diag}(\mathrm{SU}(2)_+ \times \mathrm{SU}(2)_R ) \times \mathrm{SU} (2)_-$.
\end{enumerate}
We choose $K_1'$. The supercharges transform then under $K_1'\times U(1)_R$ as
\bes
{(\bf{2},\bf{2})}^1 \oplus {(\bf{1},\bf{1})}^{-1} \oplus {(\bf{3},\bf{1})}^{-1},
\ees
The three terms combine naturally to the following operators \cite{Laba05, Ne}
\be
\label{QKL}
\begin{split}
{\CQ} &= \epsilon^{\dot{\alpha} \dot{\beta}} \bar{Q}_{\dot{\alpha} \dot{\beta}}, \\
K_{m} & =  \frac{i}{4}(\bar{\sigma}_{m})^{\dot{\alpha}\beta} Q_{\beta \dot{\alpha}},  \\
{L}_{mn} &= (\bar{\sigma}_{mn})^{\dot{\alpha}\dot{\beta}} \bar{Q}_{\dot{\alpha}\dot{\beta}}.
\end{split}
\ee
In terms of differential forms, we define $K$ and $L$ as
\begin{eqnarray*}
&K& =  K_{m}\, dx^{m} \in \Omega^1(M), \\
&L & =  L_{mn}\, dx^{m}\wedge dx^{n} \in \Omega^2(M).
\end{eqnarray*}
The $({\bf 2,2})^{1} $ representation gives thus a 1-form $K \in
\Omega^1(M)$, the $({\bf 3,1})^{-1}$ representation gives a self-dual two-form $L \in
\Omega^2(M)$, while the $({\bf 1,1})^{-1}$ representation gives $\CQ \in \Omega^0(M)$.

To determine $\{\CQ,L\}$, let us first determine the six components $\{\CQ,L_{mn}\}$.
Using the algebra (\ref{SusyAlg}) and (\ref{QKL}), we find for $(m,n)=(0,2)$ and $(1,3)$,
\begin{eqnarray*}
%\{ \bar{\CQ},\bar{L}_{12} \} &=& 0 \\
\{ {\CQ},L_{02} \} &=&  2\sqrt{2}\bar{Z},  \\
\{ {\CQ},L_{13} \} &=&  -2\sqrt{2}\bar{Z}, \
%\{ \bar{\CQ},\bar{L}_{14} \} &=&  0 \\
%\{ \bar{\CQ},\bar{L}_{23} \} &=&  0 \\
%\{ \bar{\CQ},\bar{L}_{34} \} &=&  0 \
\end{eqnarray*}
while for the other choices of $(m,n)$, $\{ \CQ, L_{mn} \}=0$. As a
result, the commutator $\{\CQ, L \}$ reads on $\BR^4$ as
\be
\{\CQ, L \} = 2\sqrt{2}\,\bar{Z}\, (dx_0 \wedge dx_{2} - dx_1 \wedge dx_3).
\ee
In complex coordinates $z_1 = x_0 + i x_2$, $z_2 = x_1 + i x_3$, we can write this commutator as follows
\be \label{QLcomplex}
\{\CQ, L \}  = \sqrt{2} i\, \bar{Z}\, \sum_{j=1,2} dz_j \wedge d\bar{z}_j \in \Omega^{1,1}(\BC^2),
\ee
We extend to an arbitrary K\"ahler surface $M$ with K\"ahler form  $J$, by realizing that $\Omega^{1,1}(M)$ contains a one-dimensional subspace of self-dual forms. Since Equation (\ref{QLcomplex}) is a $(1,1)$-form and self-dual, this suggests that
\be
\begin{split}
\{ \CQ, L \} &= \sqrt{2}i\, \bar{Z}\, J, \
\end{split}
\ee
were $J \in \Omega^{1,1}(M, \BR)$ is the K\"ahler form which spans the
one-dimensional space of $(1,1)$-forms over $M$.


\begin{thebibliography}{10}


\bibitem{Witten:1988ze}
E.~Witten, \emph{{Topological Quantum Field Theory}},
  \href{http://dx.doi.org/10.1007/BF01223371}{\emph{Commun. Math. Phys.} {\bf
  117} (1988) 353}.

\bibitem{witten1988}
E.~Witten, \emph{Topological sigma models}, {\emph{Comm. Math. Phys.} {\bf 118}
  (1988) 411--449}.

\bibitem{Kapustin:2006pk}
A.~Kapustin and E.~Witten, \emph{{Electric-Magnetic Duality And The Geometric
  Langlands Program}},
  \href{http://dx.doi.org/10.4310/CNTP.2007.v1.n1.a1}{\emph{Commun. Num. Theor.
  Phys.} {\bf 1} (2007) 1--236},
  [\href{https://arxiv.org/abs/hep-th/0604151}{{\tt hep-th/0604151}}].

\bibitem{Shapere:2008zf}
A.~D. Shapere and Y.~Tachikawa, \emph{{Central charges of N=2 superconformal
  field theories in four dimensions}},
  \href{http://dx.doi.org/10.1088/1126-6708/2008/09/109}{\emph{JHEP} {\bf 09}
  (2008) 109}, [\href{https://arxiv.org/abs/0804.1957}{{\tt 0804.1957}}].

\bibitem{Witten:1990bs}
E.~Witten, \emph{{Introduction to cohomological field theories}},
  \href{http://dx.doi.org/10.1142/S0217751X91001350}{\emph{Int. J. Mod. Phys.}
  {\bf A6} (1991) 2775--2792}.

\bibitem{Birmingham:1991ty}
M.~R. D.~Birmingham, M.~Blau and G.~Thompson, \emph{{Topological field
  theory}}, \href{http://dx.doi.org/10.1016/0370-1573(91)90117-5}{\emph{Phys.\
  Rept.\ {\bf 209}, 129 (1991)} {\bf 41} (1991) 184--244}.
 
\bibitem{Cordes:1994fc}
S.~Cordes, G.~W. Moore and S.~Ramgoolam, \emph{{Lectures on 2-d Yang-Mills
  theory, equivariant cohomology and topological field theories}},
  \href{http://dx.doi.org/10.1016/0920-5632(95)00434-B}{\emph{Nucl. Phys. Proc.
  Suppl.} {\bf 41} (1995) 184--244},
  [\href{https://arxiv.org/abs/hep-th/9411210}{{\tt hep-th/9411210}}].

\bibitem{Labastida:1991qq}
  J.~M.~F.~Labastida and P.~M.~Llatas, \emph{{Topological matter in two-dimensions}},
\href{https://doi.org/10.1016/0550-3213(92)90596-4}{\emph{Nucl.\ Phys.\ B} {\bf 379} (1992) 220--258},
  [\href{https://arxiv.org/abs/hep-th/9112051}{{\tt hep-th/9112051}}].

\bibitem{Moore:1997pc}
G.~W. Moore and E.~Witten, \emph{{Integration over the u plane in Donaldson
  theory}}, {\emph{Adv. Theor. Math. Phys.} {\bf 1} (1997) 298--387},
  [\href{https://arxiv.org/abs/hep-th/9709193}{{\tt hep-th/9709193}}].

\bibitem{LoNeSha}
A.~Losev, N.~Nekrasov and S.~L. Shatashvili, \emph{{Issues in topological gauge
  theory}}, \href{http://dx.doi.org/10.1016/S0550-3213(98)00628-2}{\emph{Nucl.
  Phys.} {\bf B534} (1998) 549--611},
  [\href{https://arxiv.org/abs/hep-th/9711108}{{\tt hep-th/9711108}}].

\bibitem{Dijkgraaf:1996tz}
R.~Dijkgraaf and G.~W. Moore, \emph{{Balanced topological field theories}},
  \href{http://dx.doi.org/10.1007/s002200050097}{\emph{Commun. Math. Phys.}
  {\bf 185} (1997) 411--440}, [\href{https://arxiv.org/abs/hep-th/9608169}{{\tt
  hep-th/9608169}}].

\bibitem{Blau:1991bn}
M.~Blau and G.~Thompson, \emph{{N=2 topological gauge theory, the Euler
  characteristic of moduli spaces, and the Casson invariant}},
  \href{http://dx.doi.org/10.1007/BF02097057}{\emph{Commun. Math. Phys.} {\bf
  152} (1993) 41--72}, [\href{https://arxiv.org/abs/hep-th/9112012}{{\tt
  hep-th/9112012}}].

\bibitem{Vafa:1994tf}
C.~Vafa and E.~Witten, \emph{{A Strong coupling test of S duality}},
  \href{http://dx.doi.org/10.1016/0550-3213(94)90097-3}{\emph{Nucl. Phys.} {\bf
  B431} (1994) 3--77}, [\href{https://arxiv.org/abs/hep-th/9408074}{{\tt
  hep-th/9408074}}].

\bibitem{DONALDSON1990257}
S.~Donaldson, \emph{Polynomial invariants for smooth four-manifolds},
  \href{http://dx.doi.org/http://dx.doi.org/10.1016/0040-9383(90)90001-Z}{\emph{Topology}
  {\bf 29} (1990) 257 -- 315}.

\bibitem{Donaldson90}
S.~K. Donaldson and P.~B. Kronheimer, \emph{The geometry of four-manifolds /
  S.K. Donaldson and P.B. Kronheimer}.
\newblock Clarendon Press ; Oxford University Press Oxford : New York, 1990.

\bibitem{Moore:2017cmm}
G.~W. Moore and I.~Nidaiev, \emph{{The Partition Function Of Argyres-Douglas
  Theory On A Four-Manifold}},  \href{https://arxiv.org/abs/1711.09257}{{\tt
  1711.09257}}.

\bibitem{Gottsche:1996aoa}
L.~Gottsche and D.~Zagier, \emph{{Jacobi forms and the structure of Donaldson
  invariants for 4-manifolds with $b_2^+=1$}}, Sel. Math., New Ser. {\bf 4} (1998) 69--115,
  \href{https://arxiv.org/abs/alg-geom/9612020}{{\tt alg-geom/9612020}}.

\bibitem{Lerche:1988np}
W.~Lerche, A.~N. Schellekens and N.~P. Warner, \emph{{Lattices and Strings}},
  \href{http://dx.doi.org/10.1016/0370-1573(89)90077-X}{\emph{Phys. Rept.} {\bf
  177} (1989) 1}.

\bibitem{Dixon:1990pc}
L.~J. Dixon, V.~Kaplunovsky and J.~Louis, \emph{{Moduli dependence of string
  loop corrections to gauge coupling constants}},
  \href{http://dx.doi.org/10.1016/0550-3213(91)90490-O}{\emph{Nucl. Phys.} {\bf
  B355} (1991) 649--688}.

\bibitem{Harvey:1995fq}
J.~A. Harvey and G.~W. Moore, \emph{{Algebras, BPS states, and strings}},
  \href{http://dx.doi.org/10.1016/0550-3213(95)00605-2}{\emph{Nucl. Phys.} {\bf
  B463} (1996) 315--368}, [\href{https://arxiv.org/abs/hep-th/9510182}{{\tt
  hep-th/9510182}}].

\bibitem{Petersson1950}
H.~Petersson, \emph{{Konstruktion der Modulformen und der zu gewissen
  Grenzkreisgruppen geh\"origen automorphen Formen von positiver reeller
  Dimension und die vollst\"andige Bestimmung ihrer Fourierkoeffizienten}},
  {\emph{S.-B. Heidelberger Akad. Wiss. Math.-Nat. Kl.} (1950) 417--494}.

\bibitem{Borcherds:1996uda}
R.~E. Borcherds, \emph{Automorphic forms with singularities on
  {G}rassmannians}, {\emph{Invent.\ Math.\ {\textbf {132}} (1998) 491
  doi:10.1007/s002220050232} (1998) },
  [\href{https://arxiv.org/abs/alg-geom/9609022}{{\tt alg-geom/9609022}}].

\bibitem{Moore:1998et}
G.~W. Moore, N.~Nekrasov and S.~Shatashvili, \emph{{D particle bound states and
  generalized instantons}},
  \href{http://dx.doi.org/10.1007/s002200050016}{\emph{Commun. Math. Phys.}
  {\bf 209} (2000) 77--95}, [\href{https://arxiv.org/abs/hep-th/9803265}{{\tt
  hep-th/9803265}}].

\bibitem{Korpas:2017qdo}
G.~Korpas and J.~Manschot, \emph{{Donaldson-Witten theory and indefinite theta
  functions}}, \href{http://dx.doi.org/10.1007/JHEP11(2017)083}{\emph{JHEP}
  {\bf 11} (2017) 083}, [\href{https://arxiv.org/abs/1707.06235}{{\tt
  1707.06235}}].

\bibitem{Korpas:2018dag}
  G.~Korpas,
``Donaldson-Witten theory, surface operators and mock modular forms,''
  arXiv:1810.07057 [hep-th].
 
\bibitem{1603.03056}
K.~Bringmann, N.~Diamantis and S.~Ehlen, \emph{Regularized inner products and
  errors of modularity},
  \href{http://dx.doi.org/10.1093/imrn/rnw225}{\emph{International Mathematics
  Research Notices} {\bf 2017} (2017) 7420--7458}.

\bibitem{bruinier2004}
J.~H. Bruinier and J.~Funke, \emph{On two geometric theta lifts},
  \href{http://dx.doi.org/10.1215/S0012-7094-04-12513-8}{\emph{Duke Math. J.}
  {\bf 125} (10, 2004) 45--90}.

\bibitem{Duke2016}
W.~Duke, {\"O}.~Imamo\={g}lu and {\'A}.~T{\'o}th, \emph{Regularized inner
  products of modular functions},
  \href{http://dx.doi.org/10.1007/s11139-013-9544-5}{\emph{The Ramanujan
  Journal} {\bf 41} (Nov, 2016) 13--29}.

\bibitem{Seiberg:1994rs}
N.~Seiberg and E.~Witten, \emph{{Electric - magnetic duality, monopole
  condensation, and confinement in N=2 supersymmetric Yang-Mills theory}},
  \href{http://dx.doi.org/10.1016/0550-3213(94)90124-4,
  10.1016/0550-3213(94)00449-8}{\emph{Nucl. Phys.} {\bf B426} (1994) 19--52},
  [\href{https://arxiv.org/abs/hep-th/9407087}{{\tt hep-th/9407087}}].

\bibitem{Laba05}
J.~Labastida and M.~Marino, \emph{{Topological quantum field theory and four
  manifolds}}, vol.~25.
\newblock Springer, Dordrecht, 2005,
  \href{http://dx.doi.org/10.1007/1-4020-3177-7}{10.1007/1-4020-3177-7}.

\bibitem{MooreNotes2017}
G.~W. Moore, \emph{{Lectures On The Physical Approach To Donaldson And
  Seiberg-Witten Invariants}}, {\emph{Item 78 at
  http://www.physics.rutgers.edu/~gmoore/} (2017) }.

\bibitem{Ne}
N.~A. Nekrasov, \emph{{Seiberg-Witten prepotential from instanton counting}},
  \href{http://dx.doi.org/10.4310/ATMP.2003.v7.n5.a4}{\emph{Adv. Theor. Math.
  Phys.} {\bf 7} (2003) 831--864},
  [\href{https://arxiv.org/abs/hep-th/0206161}{{\tt hep-th/0206161}}].

\bibitem{Witten:1995gf}
E.~Witten, \emph{{On S duality in Abelian gauge theory}},
  \href{http://dx.doi.org/10.1007/BF01671570}{\emph{Selecta Math.} {\bf 1}
  (1995) 383}, [\href{https://arxiv.org/abs/hep-th/9505186}{{\tt
  hep-th/9505186}}].

\bibitem{Matone:1995rx}
M.~Matone, \emph{{Instantons and recursion relations in N=2 SUSY gauge
  theory}}, \href{http://dx.doi.org/10.1016/0370-2693(95)00920-G}{\emph{Phys.
  Lett.} {\bf B357} (1995) 342--348},
  [\href{https://arxiv.org/abs/hep-th/9506102}{{\tt hep-th/9506102}}].

\bibitem{to_appear}
G.~Korpas, J.~Manschot, G.~W. Moore and I.~Nidaiev, \emph{To appear}, .

\bibitem{Vigneras:1977}
M.-F. Vign\'eras, \emph{{S\'eries th\^eta des formes quadratiques
  ind\'efinies}}, {\emph{Springer Lecture Notes} {\bf 627} (1977) 227 -- 239}.
 
\bibitem{Larry}
K.~Bringmann, A.~Folsom, K.~Ono and L.~Rolen, \emph{Harmonic {M}aass forms and
  mock modular forms: theory and applications}, vol.~64 of \emph{American
  Mathematical Society Colloquium Publications}.
\newblock American Mathematical Society, Providence, RI, 2017.

\bibitem{ZwegersThesis}
S.~P. Zwegers, \emph{Mock Theta Functions}.
\newblock PhD thesis, 2008.

\bibitem{MR2605321}
D.~Zagier, \emph{Ramanujan's mock theta functions and their applications (after
  {Z}wegers and {O}no-{B}ringmann)}, {\emph{Ast\'erisque} (2009) Exp. No. 986,
  vii--viii, 143--164 (2010)}.

\bibitem{Polchinski:1998rq}
J.~Polchinski, \emph{{String theory. Vol. 1: An introduction to the bosonic
  string}}.
\newblock Cambridge Monographs on Mathematical Physics. Cambridge University
  Press, 2007,
  \href{http://dx.doi.org/10.1017/CBO9780511816079}{10.1017/CBO9780511816079}.

\bibitem{Eberhardt:2023xck}
L.~Eberhardt and S.~Mizera,
\emph{Evaluating one-loop string amplitudes}, \href{https://arxiv.org/abs/2302.12733}{{\tt
  2302.12733}}.

\bibitem{Witten:2013pra}
E.~Witten,
\emph{The Feynman $i \epsilon$ in String Theory,}
JHEP \textbf{04} (2015), 055
doi:10.1007/JHEP04(2015)055 \href{https://arxiv.org/abs/1307.5124}{{\tt
  1307.5124}}.
  
\bibitem{Serre}
J.~P. Serre, \emph{A course in arithmetic}.
\newblock Graduate Texts in Mathematics, no. 7, Springer, New York, 1973.

\bibitem{Zagier92}
D.~Zagier, \emph{Introduction to modular forms; From Number Theory to Physics}.
\newblock Springer, Berlin (1992), pp. 238-291, 1992.

\bibitem{Bruinier08}
G.~H. J.H.~Bruinier, G. van der~Geer and D.~Zagier, \emph{The 1-2-3 of Modular
  Forms}.
\newblock Springer-Verlag Berlin Heidelberg, 2008,
  \href{http://dx.doi.org/10.1007/978-3-540-74119-0}{10.1007/978-3-540-74119-0}.


\end{thebibliography}
\end{document}